\documentclass[useAMS,usenatbib]{mn2e}

\usepackage{txfonts}
\usepackage{graphicx}
\usepackage{subfigure}

\title[Relative motions of dense cores and envelopes]{On the relative motions of dense cores and envelopes in star-forming molecular clouds}
\author[B.A. Ayliffe et al.]{Ben A. Ayliffe, James C. Langdon, Howard S. Cohl, and Matthew R. Bate\thanks{E-mail:
mbate@astro.ex.ac.uk}\\ School of Physics, University of Exeter, Stocker
Road, Exeter EX4 4QL}

\date{Accepted 2006 October 28. Received 2006 October 3; in original form 2006 March 13}
\begin{document}
\maketitle
\begin{abstract}
Hydrodynamical simulations of star formation indicate that the motions of protostars through their natal molecular clouds may be crucial in determining the properties of stars through competitive accretion and dynamical interactions.  Walsh, Myers \& Burton recently investigated whether such motions might be observable in the earliest stages of star formation by measuring the relative shifts of line-centre velocities of low- and high-density tracers of low-mass star-forming cores.  They found very small ($\sim 0.1$ km~s$^{-1}$) relative motions.  In this paper, we analyse the hydrodynamical simulation of Bate, Bonnell \& Bromm and find that it also gives small relative velocities between high-density cores and low-density envelopes, despite the fact that competitive accretion and dynamical interactions occur between protostars in the simulation.  Thus, the simulation is consistent with the observations in this respect.  However, we also find some differences between the simulation and the observations.  Overall, we find that the high-density gas has a higher velocity dispersion than that observed by Walsh et al.  We  explore this by examining the dependence of the gas velocity dispersion on density and its evolution with time during the simulation.  We find that early in the simulation the gas velocity dispersion decreases monotonically with increasing density, while later in the simulation, when the dense cores have formed multiple objects, the velocity dispersion of the high-density gas increases.  Thus, the simulation is in best agreement with the observations early on, before many objects have formed in each dense core.
\end{abstract}

\begin{figure*}
\centering
\subfigure 
{
    \includegraphics[width=4.2cm]{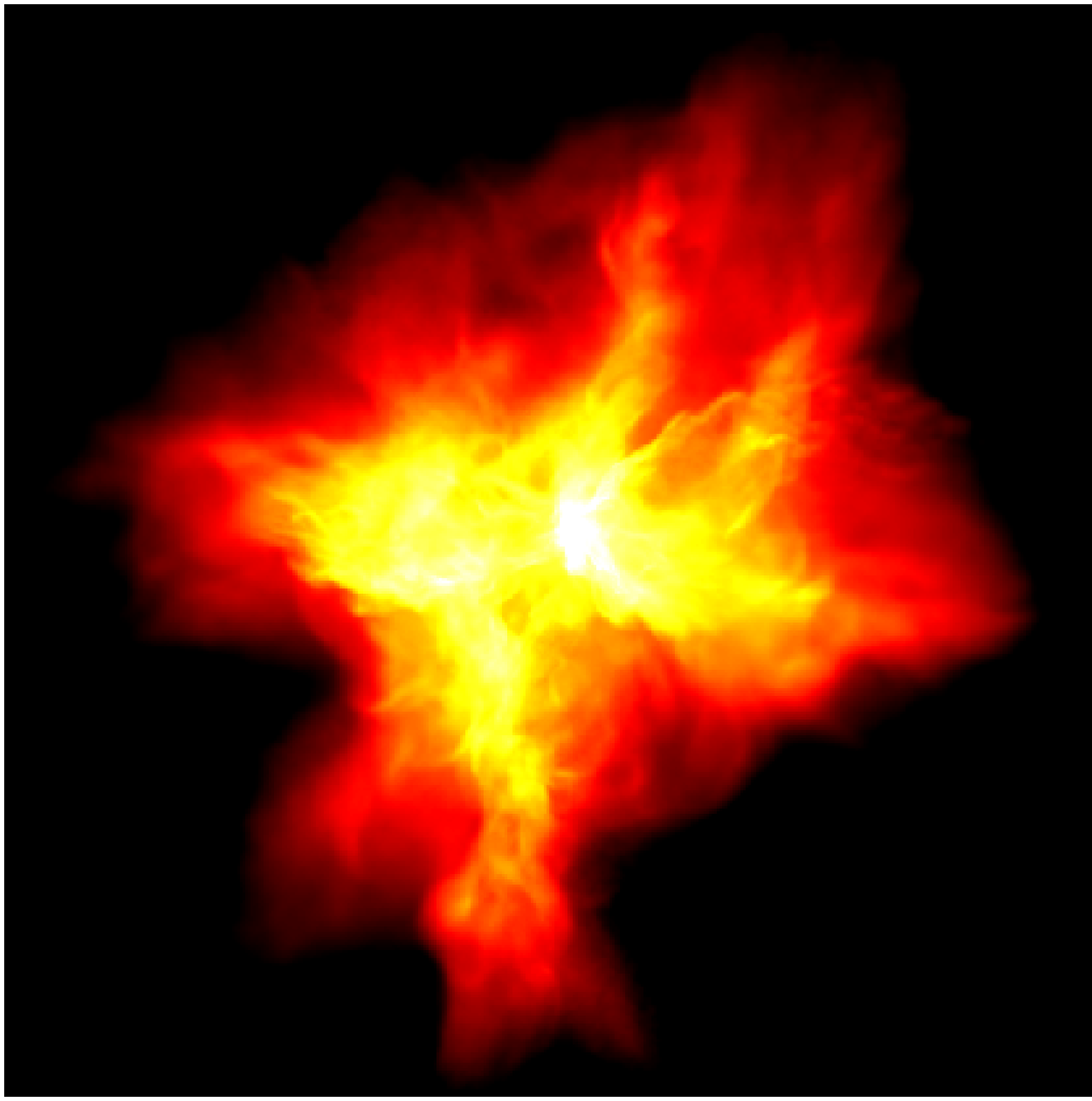}
}
\subfigure 
{
    \includegraphics[width=4.2cm]{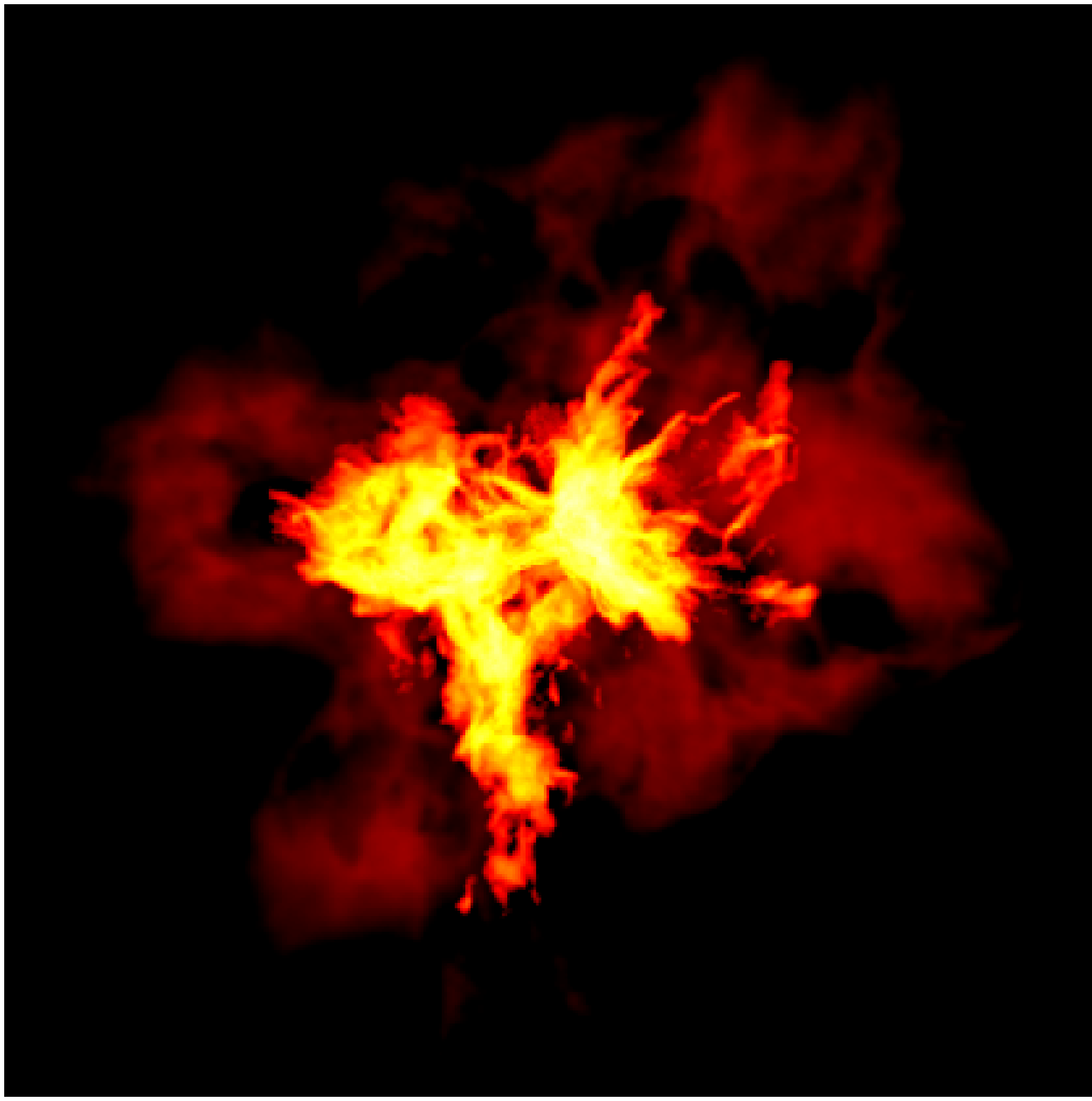}
}
\subfigure 
{
    \includegraphics[width=4.2cm]{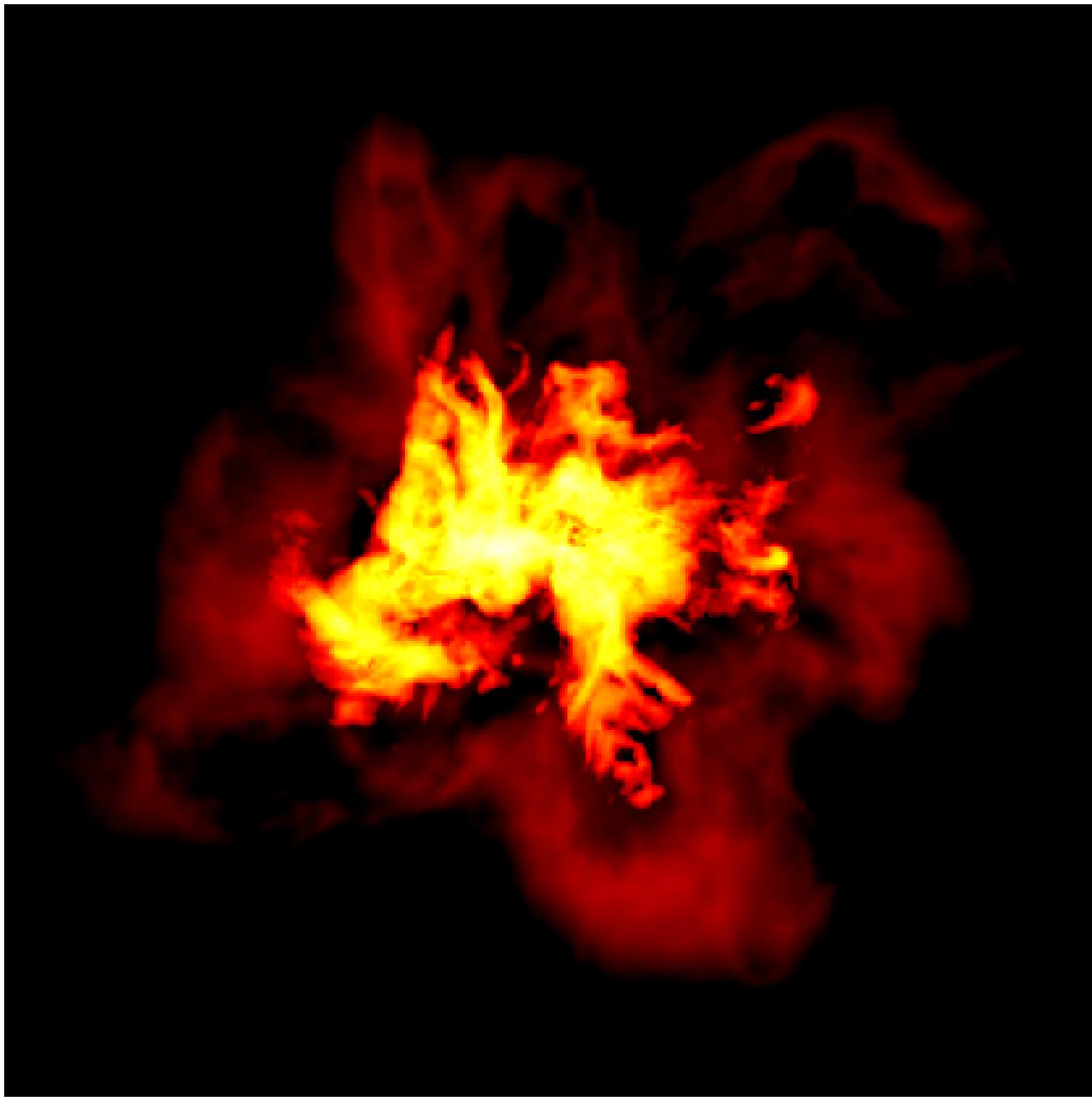}
}

\vspace{-0.2cm}

\subfigure 
{
    \includegraphics[width=4.2cm]{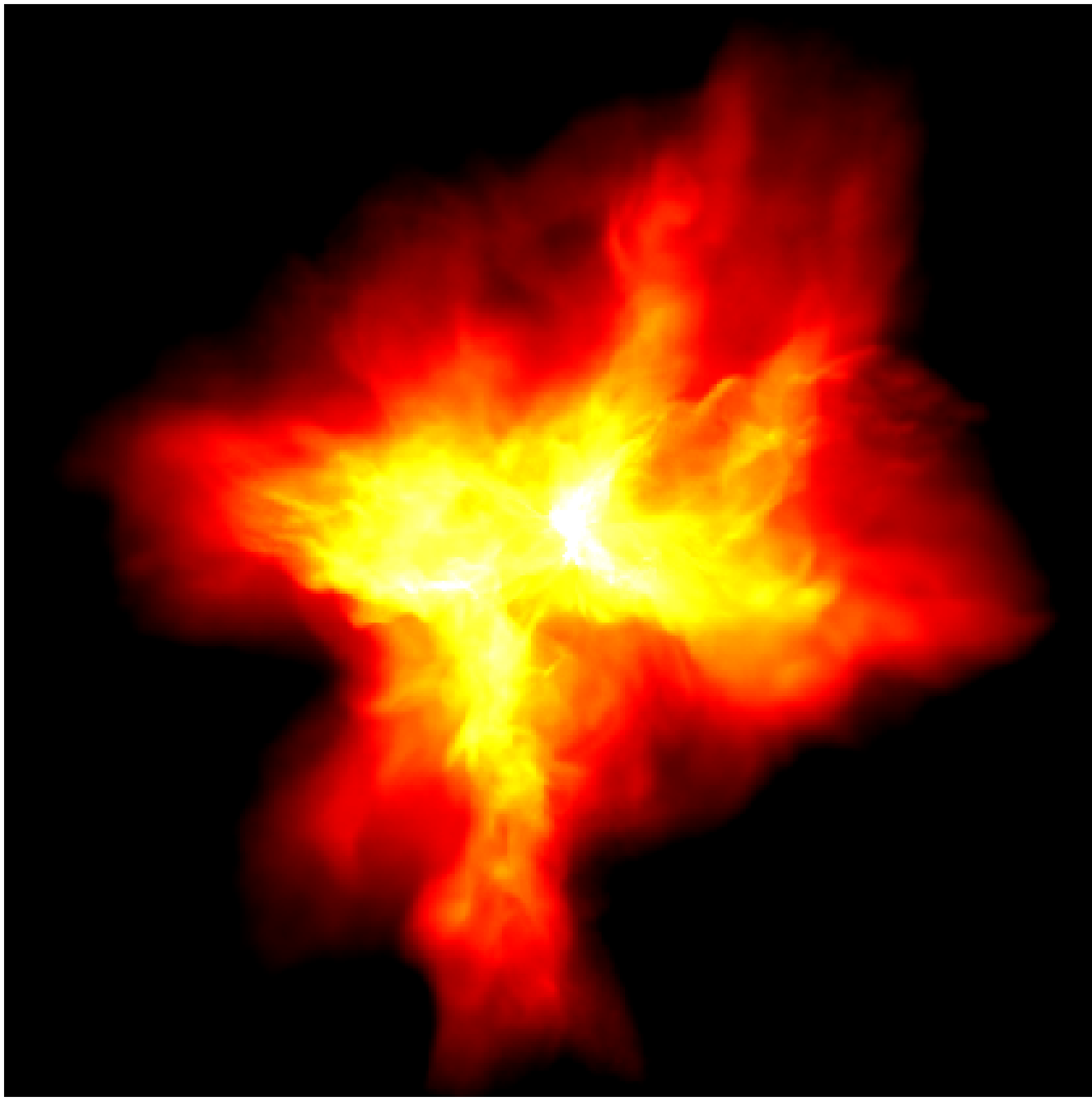}
}
\subfigure 
{
    \includegraphics[width=4.2cm]{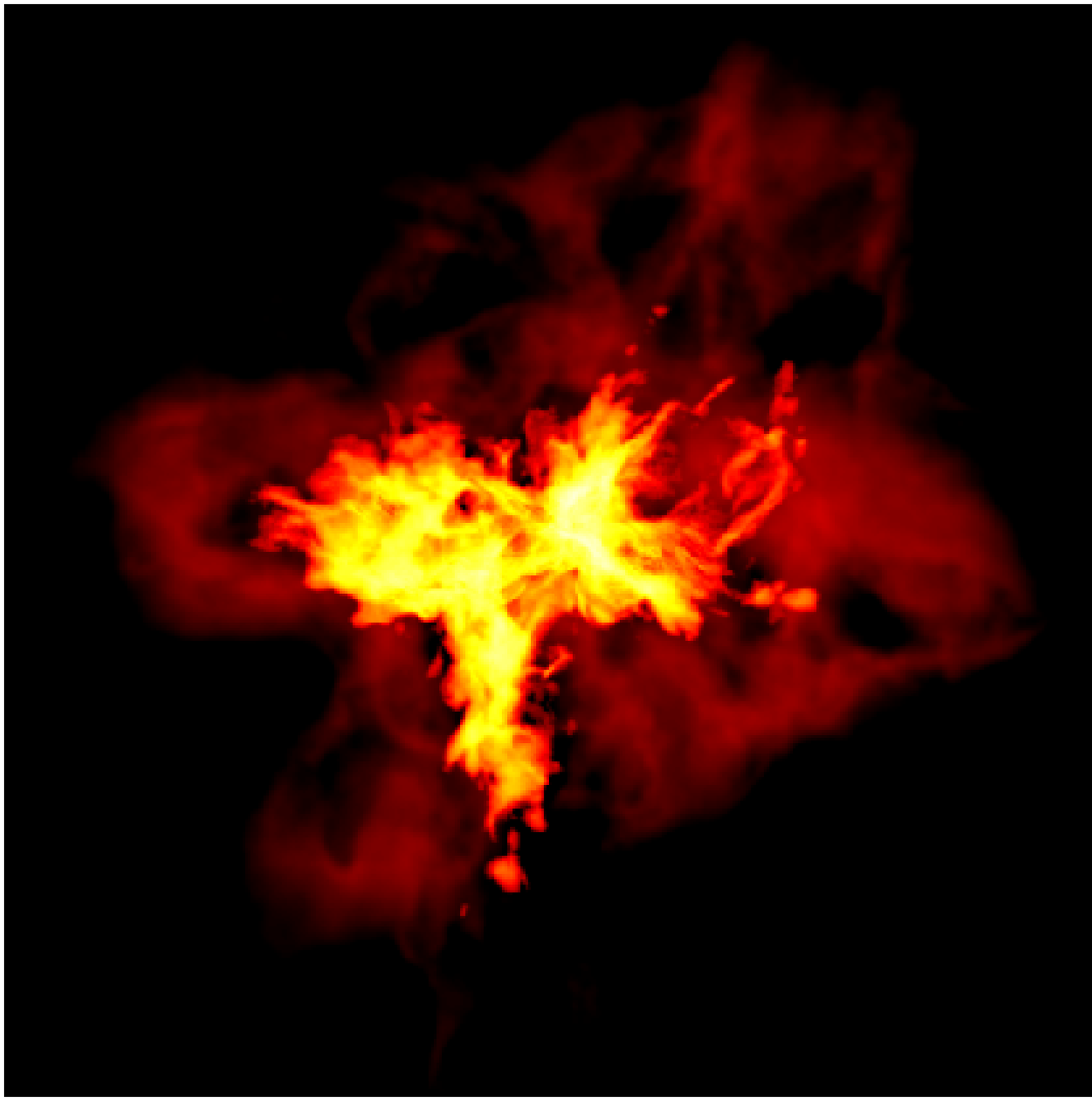}
}
\subfigure 
{
    \includegraphics[width=4.2cm]{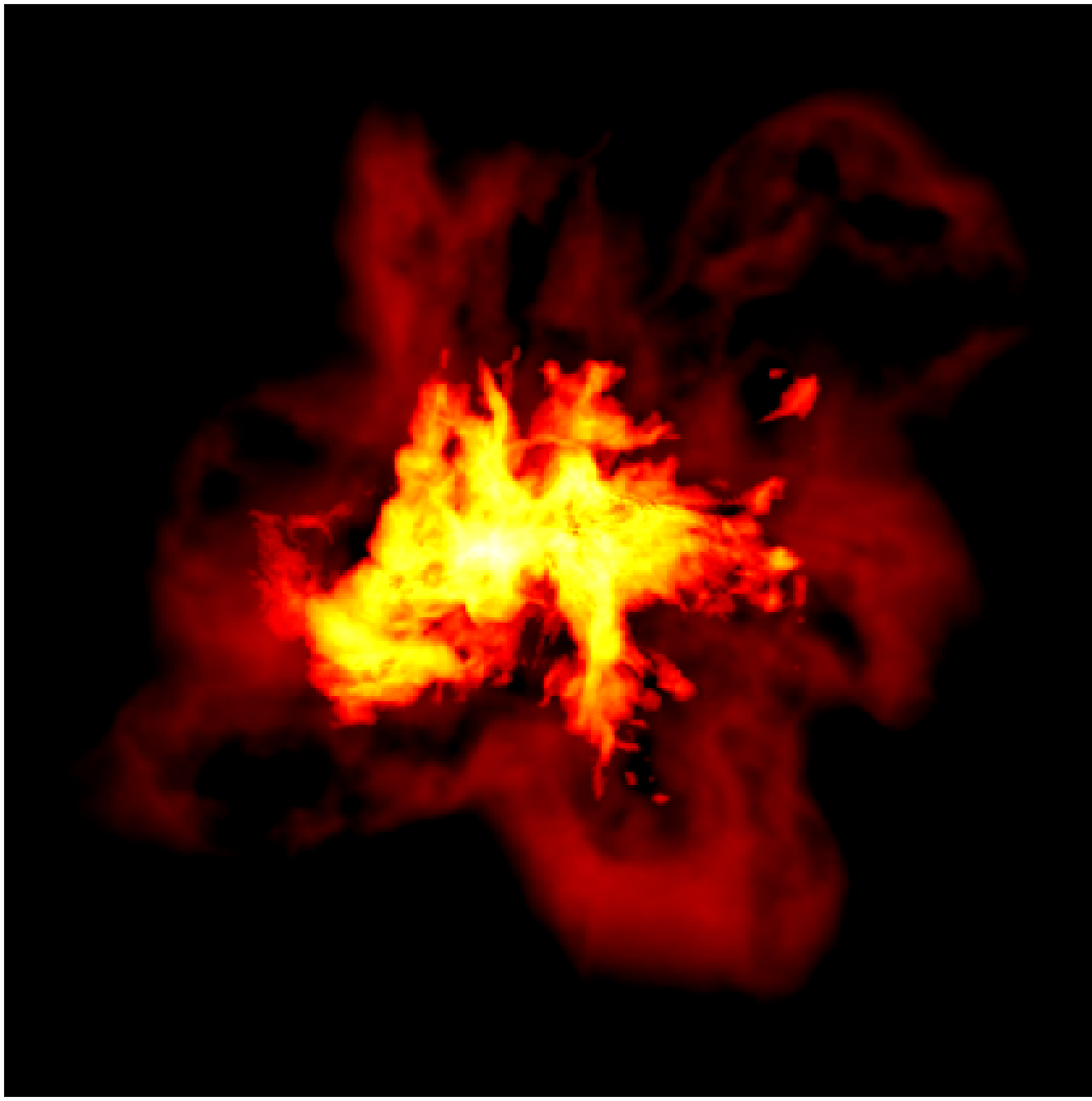}
}

\vspace{-0.2cm}

\subfigure 
{
    \includegraphics[width=4.2cm]{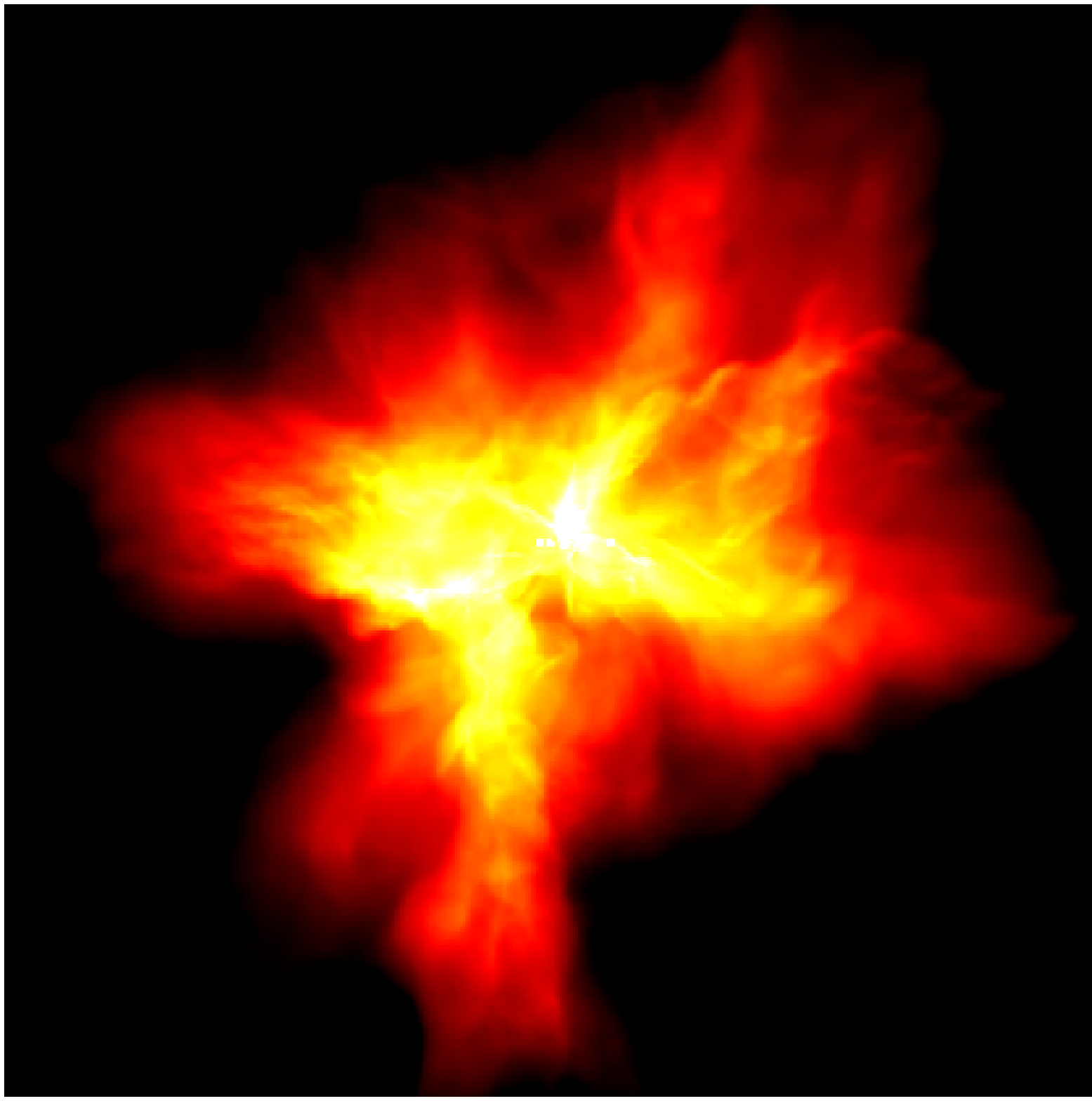}
}
\subfigure 
{
    \includegraphics[width=4.2cm]{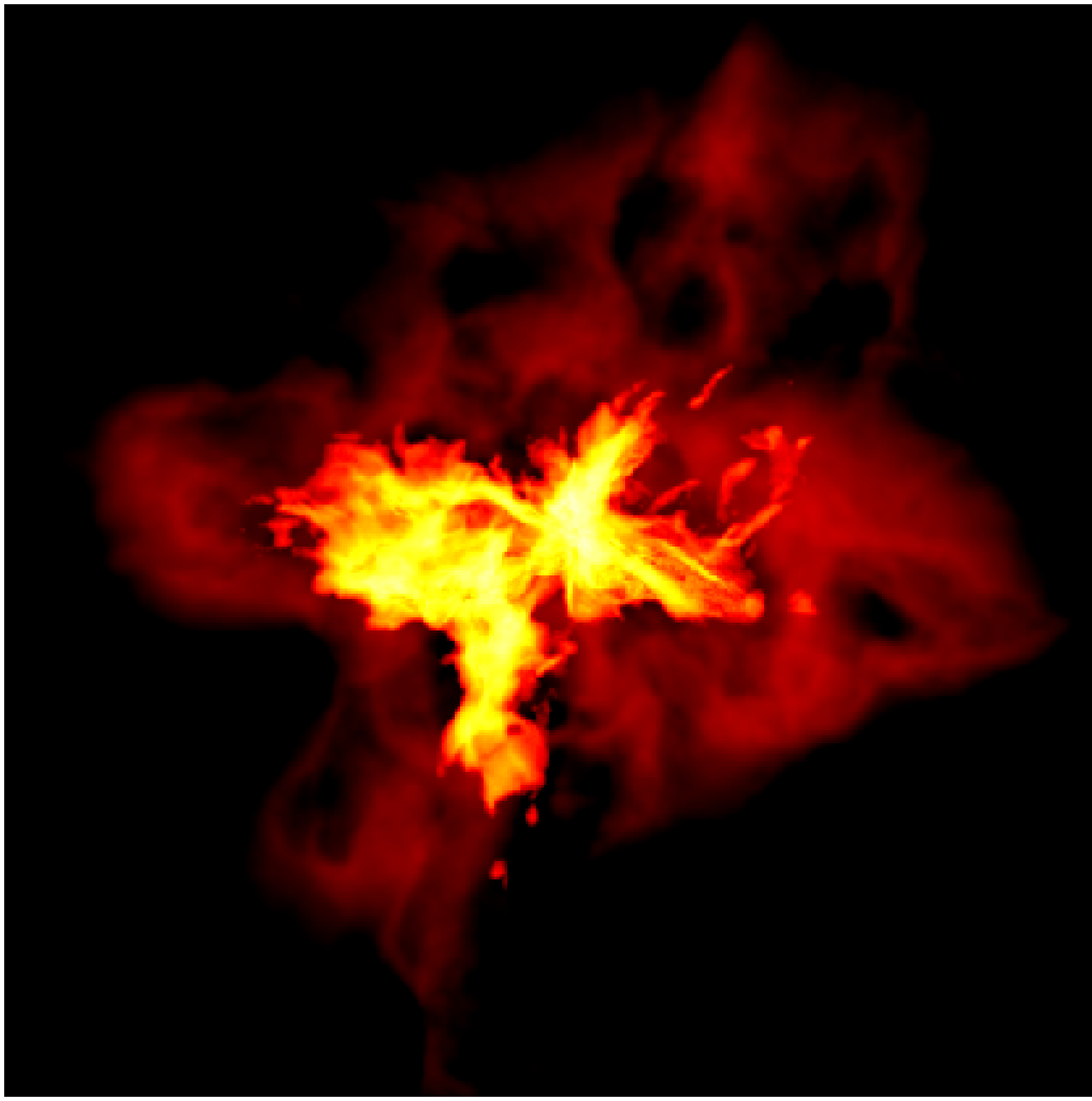}
}
\subfigure 
{
    \includegraphics[width=4.2cm]{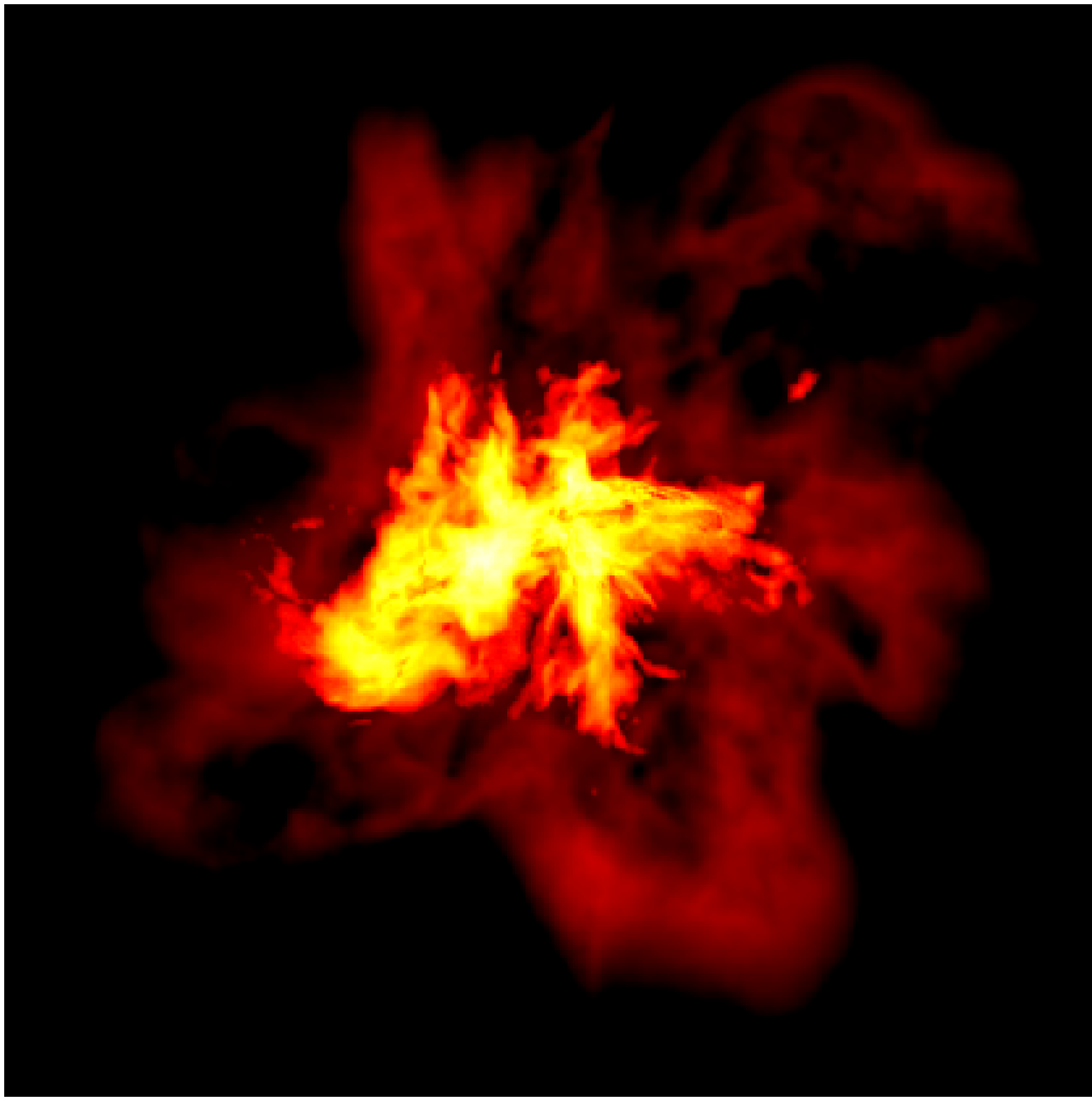}
}

\vspace{-0.2cm}

\subfigure 
{
    \includegraphics[width=4.2cm]{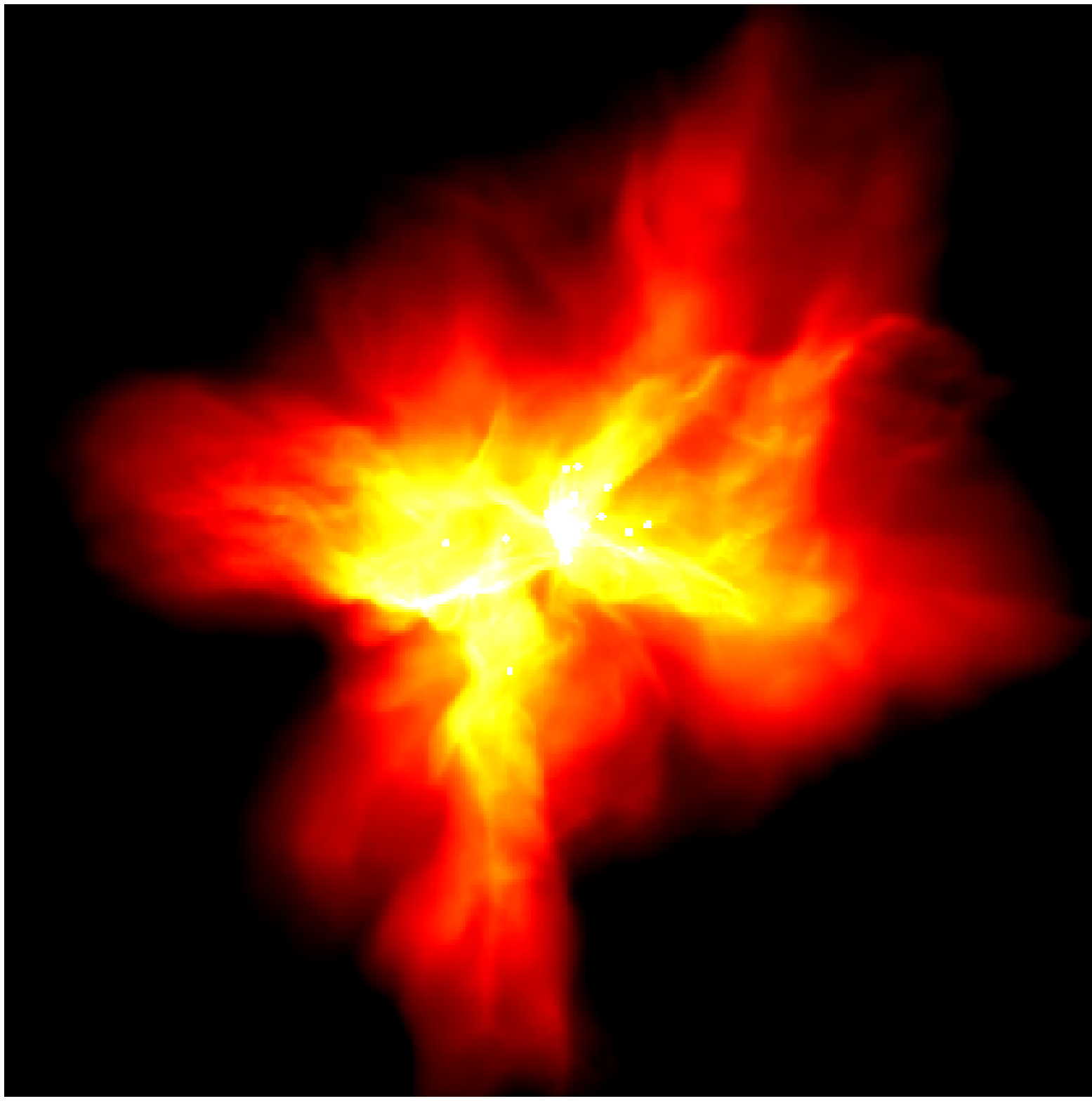}
}
\subfigure 
{
    \includegraphics[width=4.2cm]{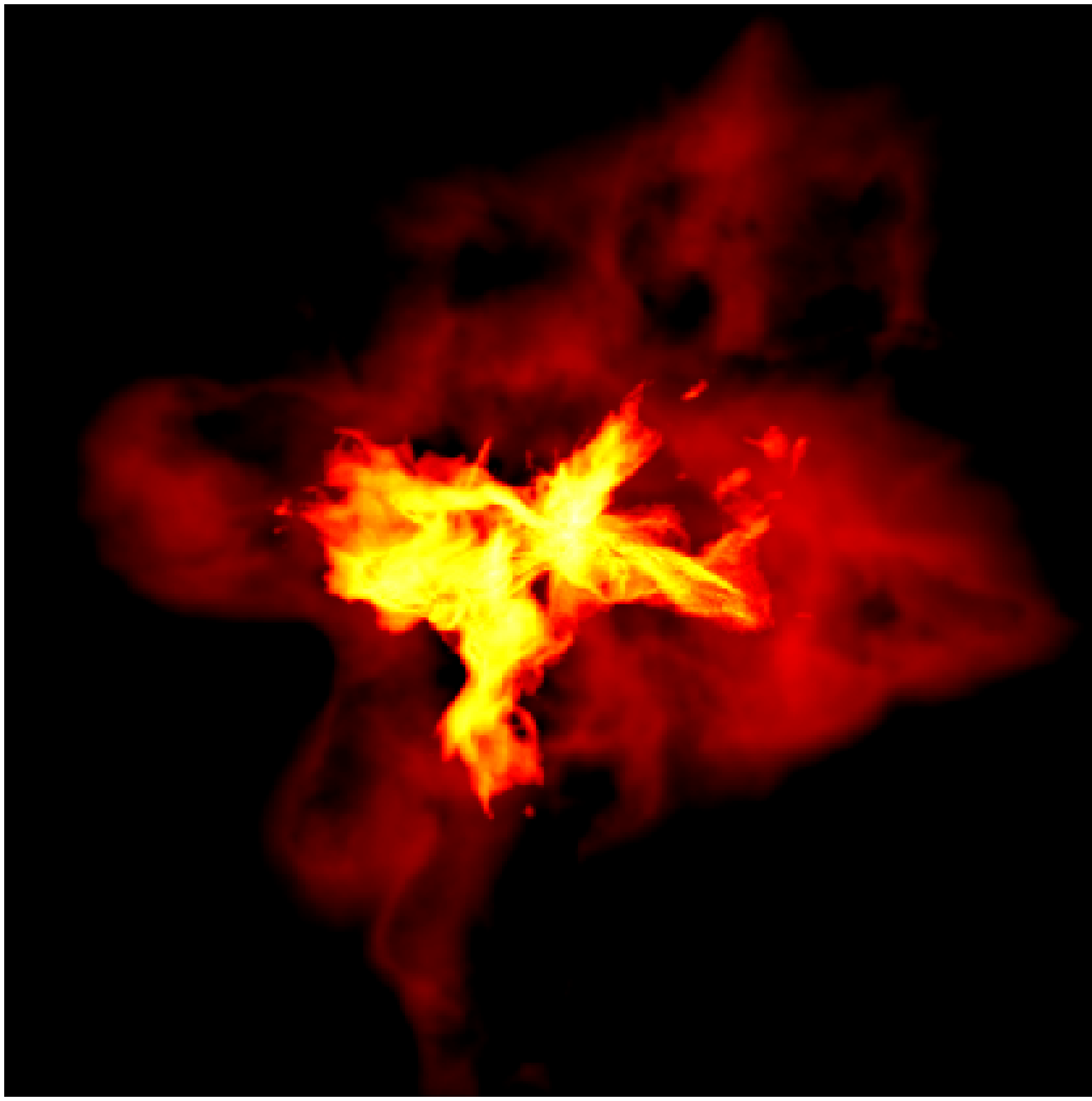}
}
\subfigure 
{
    \includegraphics[width=4.2cm]{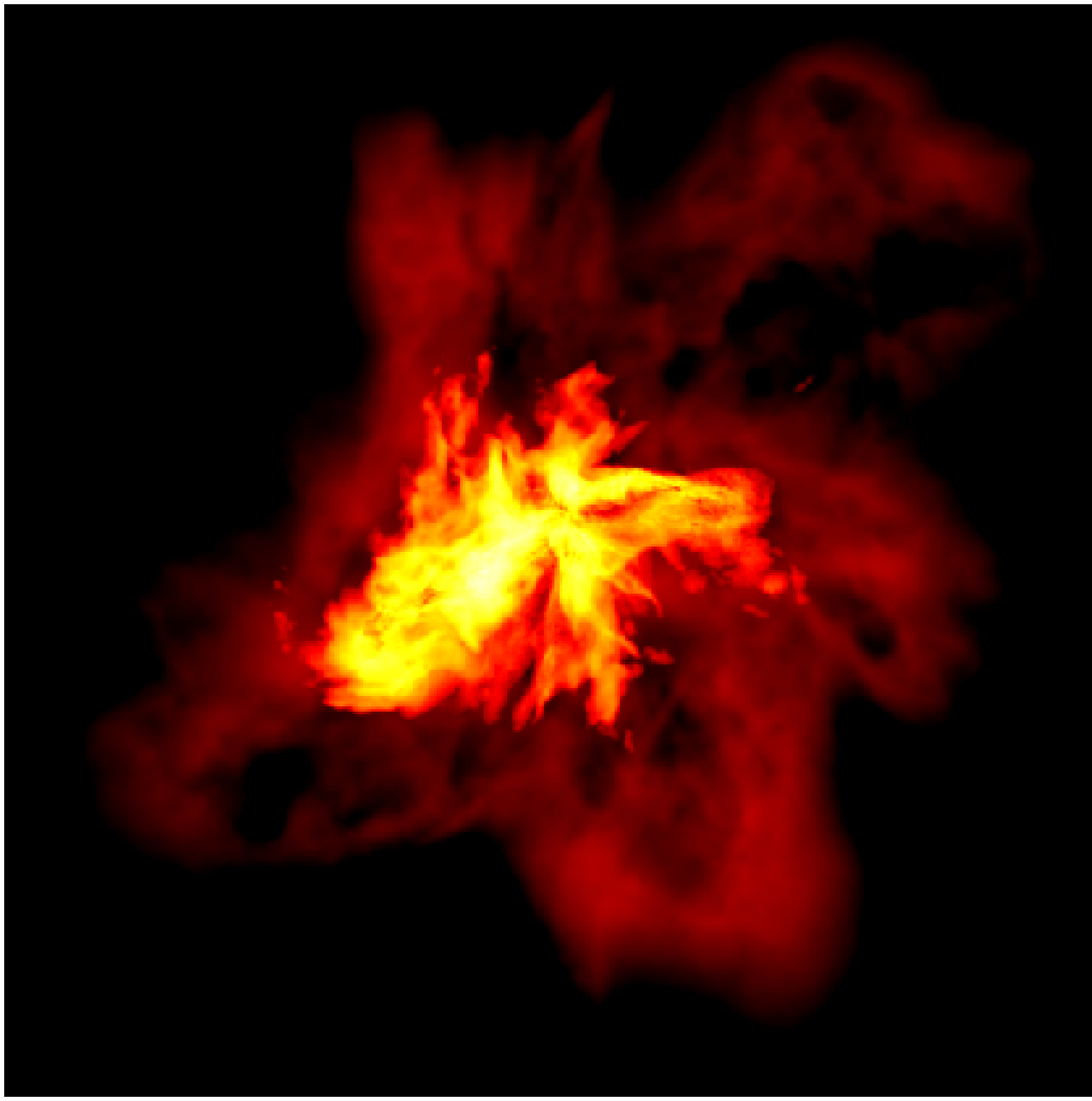}
}

\vspace{-0.2cm}

\subfigure 
{
    \includegraphics[width=4.2cm]{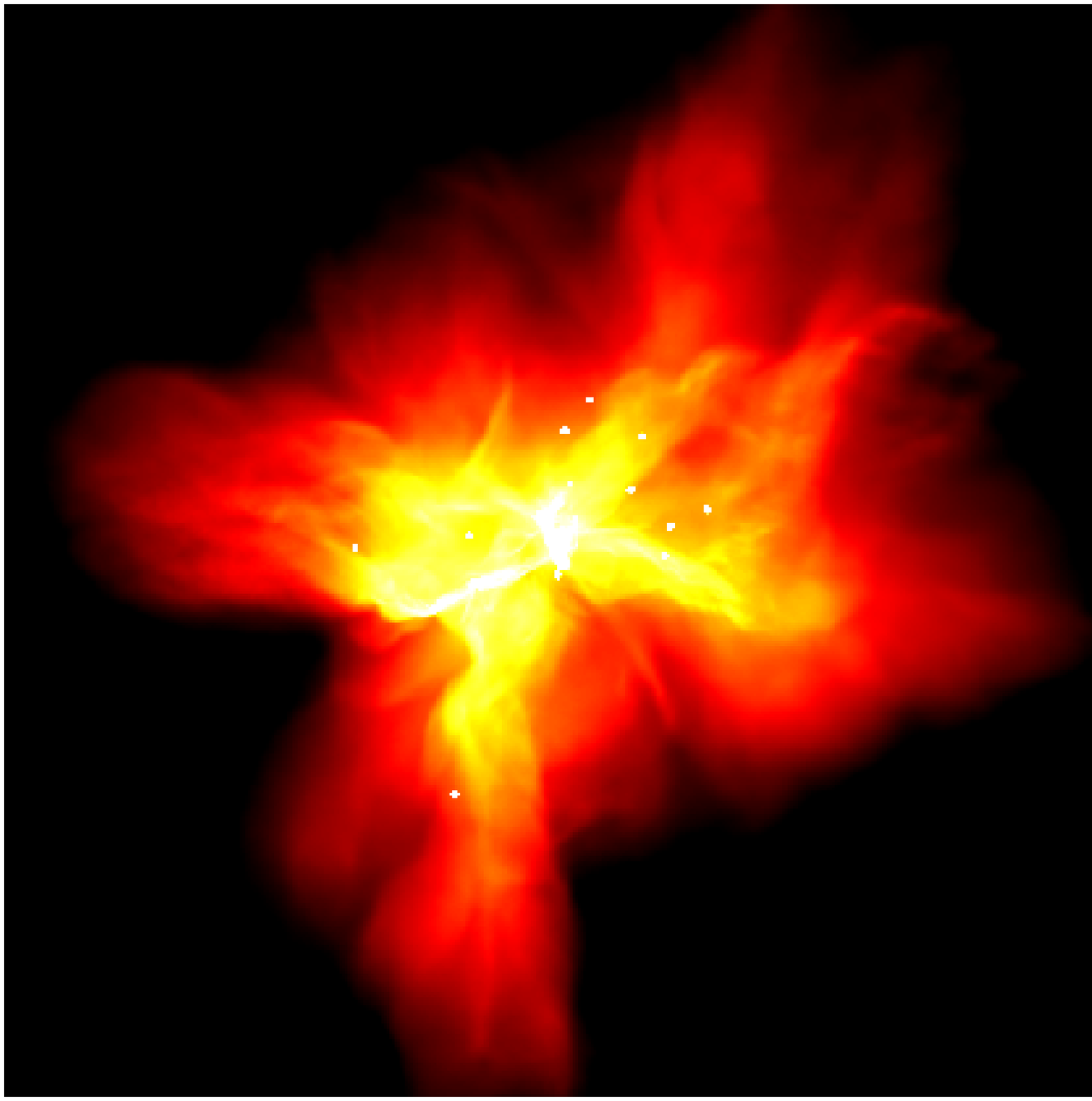}
}
\subfigure 
{
    \includegraphics[width=4.2cm]{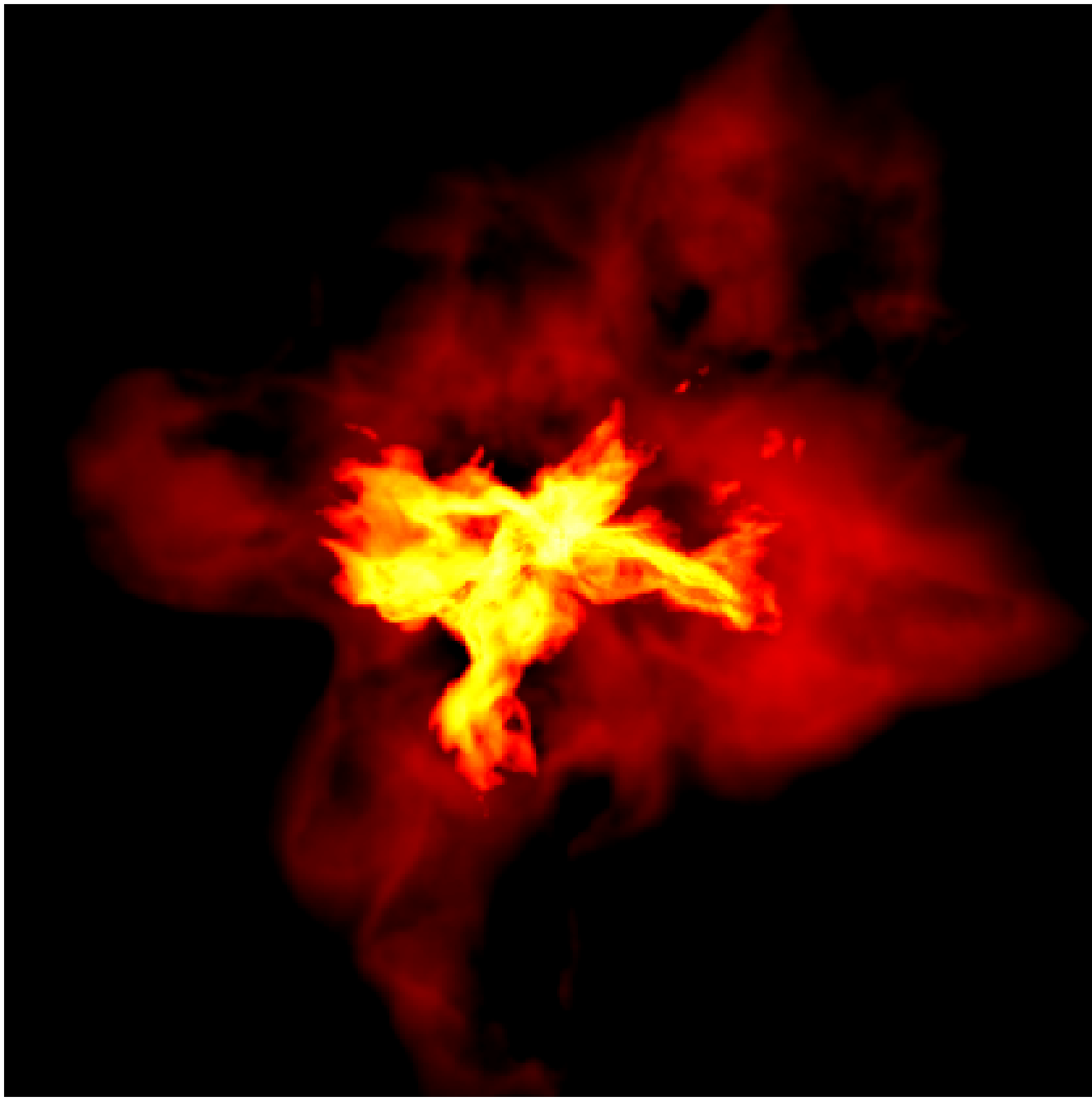}
}
\subfigure 
{
    \includegraphics[width=4.2cm]{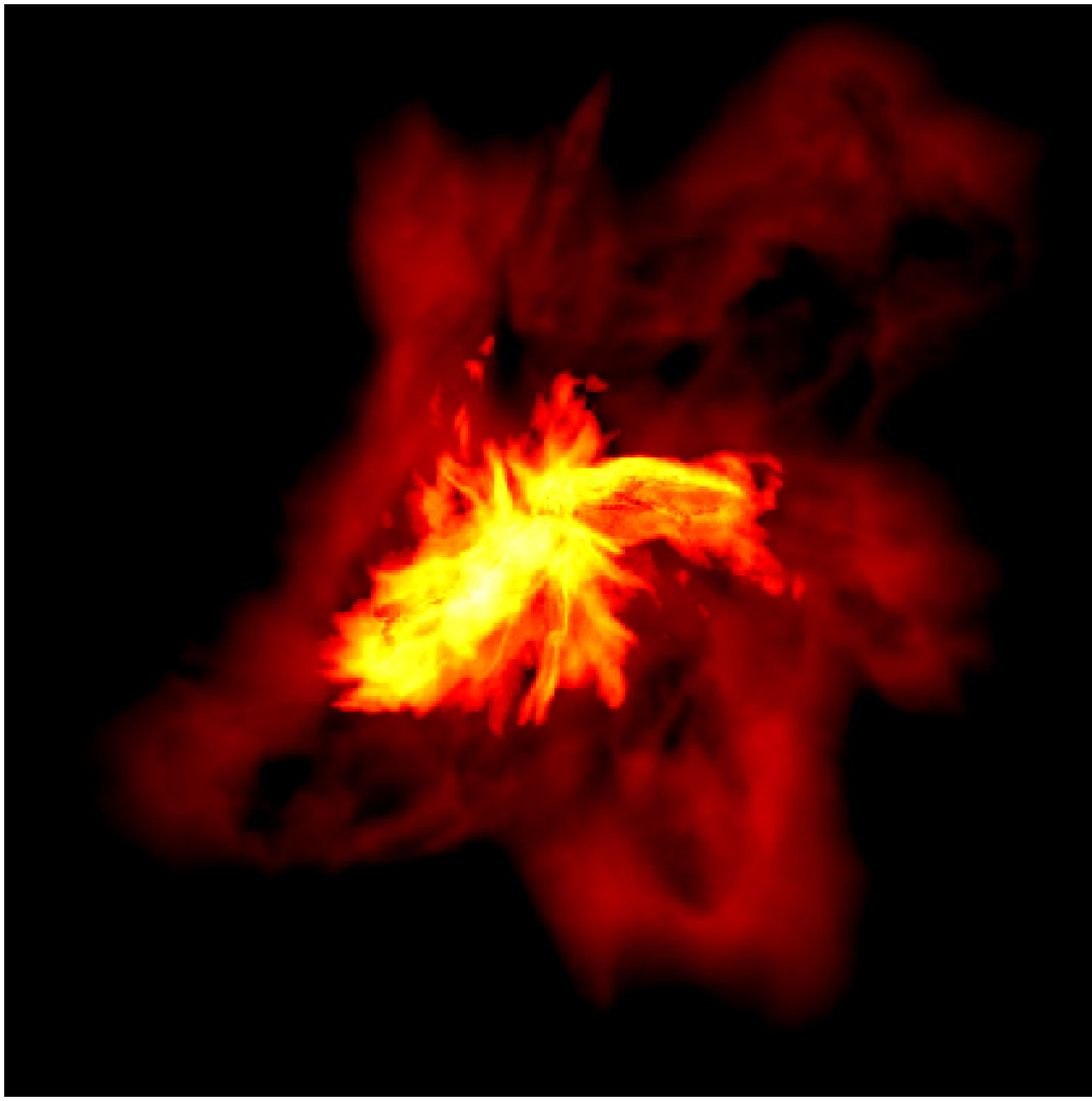}
}

\caption{The star-forming cloud at five different times ($t=1.0, 1.1, 1.2, 1.3,$ and 1.4 $t_{\rm ff}$, top to bottom).  At each time, we plot the full column-density maps (left panels, see also Bate et al.\ 2003) while in the centre and right panels we plot only the high-density and low-density gas components as defined in the main text (centre panels from the same angle as the left panels, right panels by rotated $90^\circ$ perpendicular to the page).  The panels measure 0.8 pc across, and the column density ranges from $\log N = -2.7$ to $-0.3$ 	with $N$ measured in g~cm$^{-2}$.}
\label{pics}
\end{figure*}

\section{Introduction}

Over the past eight years, hydrodynamical simulations of clustered star formation have led to the realisation that dynamical processes may be crucial to determining the statistical properties of stars.  As opposed to the isolated star formation model \citep*[e.g.,][]{ShuAdaLiz1987, PadNor2002}, where a star's final mass is determined by the mass of the dense core that is accreted on to the star, protostars in a cluster move around within the overall gravitational potential well of the gas-rich cluster accreting competitively (\citealt{Zinnecker1982, Bonnelletal1997, Bonnelletal2001a}; \citealt*{BonBatVin2003}).   The most massive stars form near the centre of the cluster into which gas is funneled by the overall gravitational potential well and where the gas densities, and therefore the accretion rates, are highest.  This may explain why even very young clusters are mass segregated \citep{BonDav1998, HilHar1998}.  Simultaneously, dynamical interactions between protostars may eject objects from the gas-rich dense cores, terminating their accretion and setting their final masses \citep*{KleBurBat1998}.  In particular, such dynamical ejections may be responsible for producing low-mass stars and brown dwarfs \citep*{ReiCla2001, BatBonBro2002a}.  Along with the stellar initial mass function (IMF), dynamical interactions may also be crucial for the production of close binary systems \citep*{BatBonBro2002b}, providing the velocity dispersion in star-forming regions \citep*{BatBonBro2003, BatBon2005}, and truncating circumstellar discs \citep{BatBonBro2003}.

Recently, \citet*{WalMyeBur2004} investigated whether or not such dynamical motions arise very early in the star formation process.  Specifically, they examined whether dense cores that are likely to collapse to form stars or that are already forming stars are moving ballistically relative to their low-density envelopes.  They used N$_2$H$^+$ ($1-0$) observations to identify high-density cores and measure their line-centre velocities and $^{13}$CO ($1-0$) or C$^{18}$O ($1-0$) observations to find their corresponding low-density envelopes and measure their line-centre velocities.  If dense cores move ballistically, they expected to find differences in the line-center velocities comparable to the line width of the CO spectra, assuming that the CO line width is representative of the escape velocity of the envelope.  Alternatively, if dense cores do not move ballistically, they expected to see much smaller differences in line-center velocities, similar to or smaller than the N$_2$H$^+$ line width, which traces the internal random motions within dense cores.  They found that the distribution of line-centre velocity differences was certainly narrower than the average $^{13}$CO and C$^{18}$O line widths and was even somewhat narrower than the N$_2$H$^+$ line width.  Thus, they concluded that the dense cores did not move ballistically with respect to their envelopes.  This implies that dense cores do not gain significant mass by accreting gas as they move through the lower-density environment.

An obvious question raised by Walsh et al.'s study is whether hydrodynamical simulations in which competitive accretion and dynamical interactions between protostars lead to the statistical properties of stars are consistent with their observations or not.  In this paper, we analyse the simulation of \citet{BatBonBro2003} in a similar manner to Walsh et al.\ to determine whether dense cores in the simulation have large or small velocities relative to the low-density envelopes in which they are embedded.  Bate et al.\ performed a hydrodynamical simulation of the collapse of a 50 solar mass molecular cloud to form a small cluster of 50 stars and brown dwarfs.  We find that, in agreement with the observations, there are only small line-centre velocity differences between high-density cores and their associated low-density envelopes.  In Section \ref{method}, we describe the method we used to analyse the simulation and compare it to that of Walsh et al.  In Section \ref{results}, we present our results.  Our conclusions are given in Section \ref{conclusions}.

\begin{figure*}
\centering
{
    \label{contour:sub:a}
    \includegraphics[width=7.1cm]{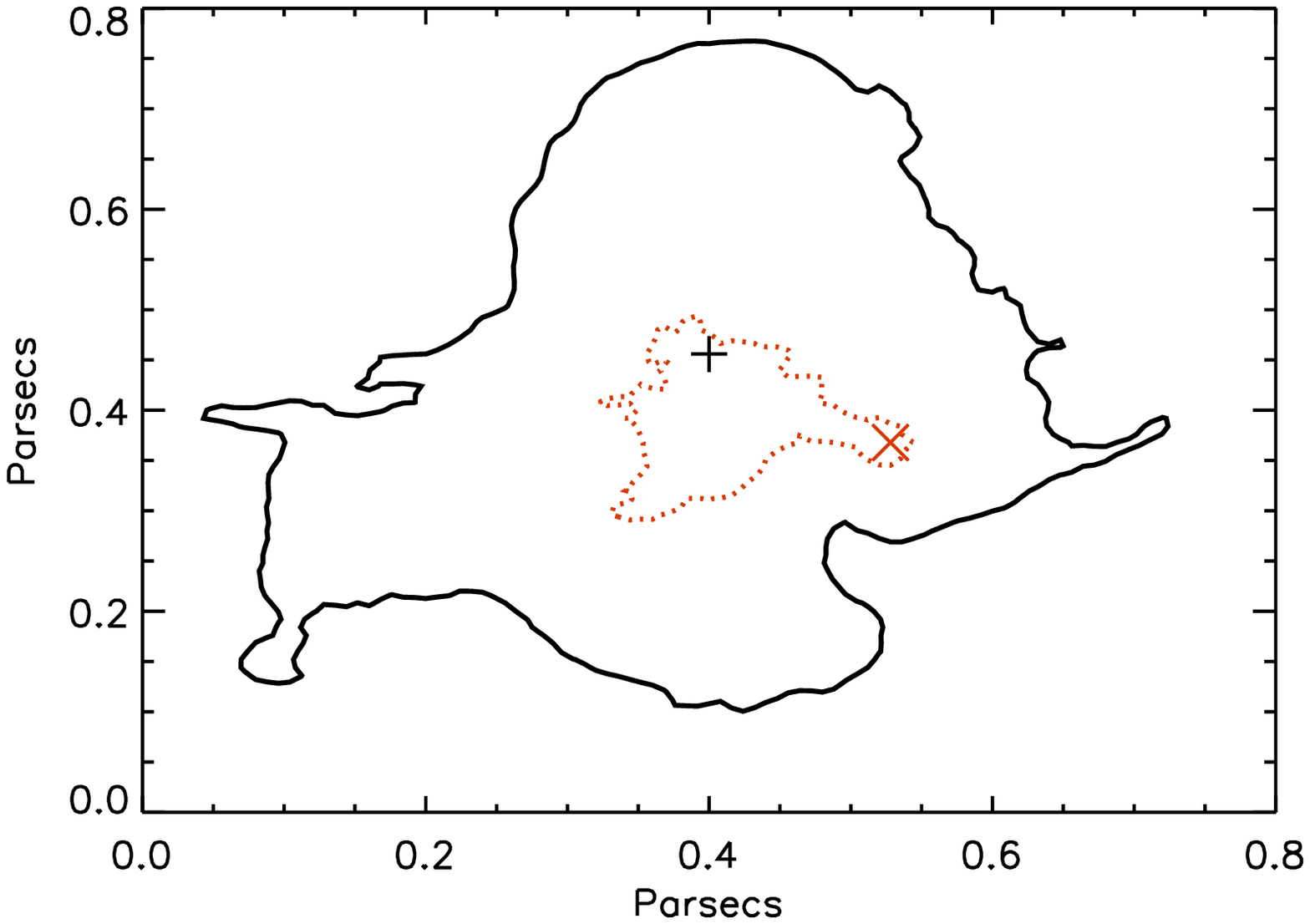}
}
{
    \label{contour:sub:b}
    \includegraphics[width=7.0cm]{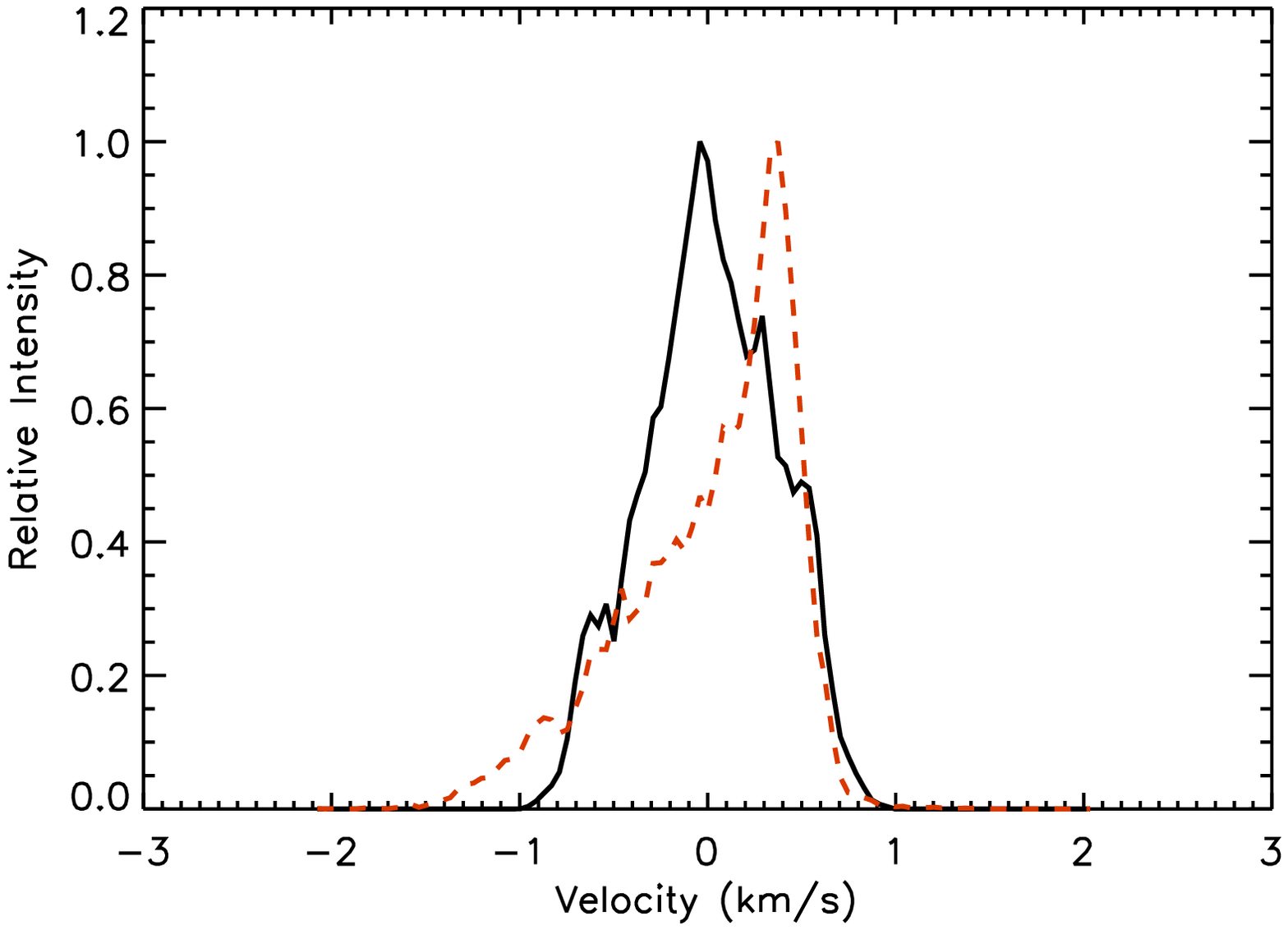}
}
\caption{\label{example} An example (left panel) of the identification from the simulation of a high-density core (red dotted line) and its low-density envelope (black solid line) and their velocity spectra (right panel).  The contours show the regions within which the column densities are greater than 50\% of the core ($\times$) and envelope (+) peak values.  The velocity spectra are constructed only from gas that lies within the high-density contour, as in Walsh et al.}
\label{contour}
\end{figure*}

\begin{figure*}
\centering
{
    \label{contoursoft:sub:a}
    \includegraphics[width=7.1cm]{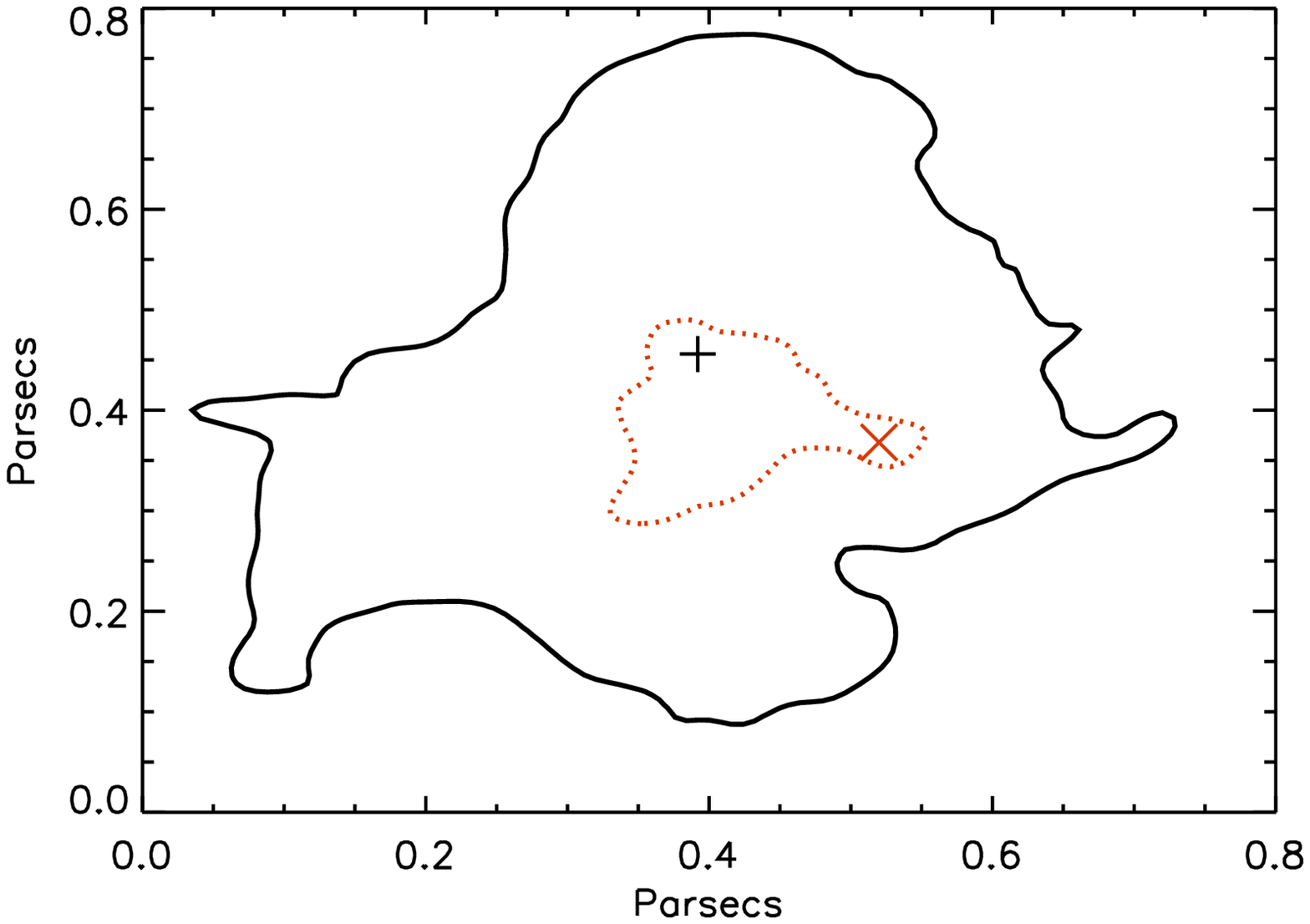}
}
{
    \label{contoursoft:sub:b}
    \includegraphics[width=7.0cm]{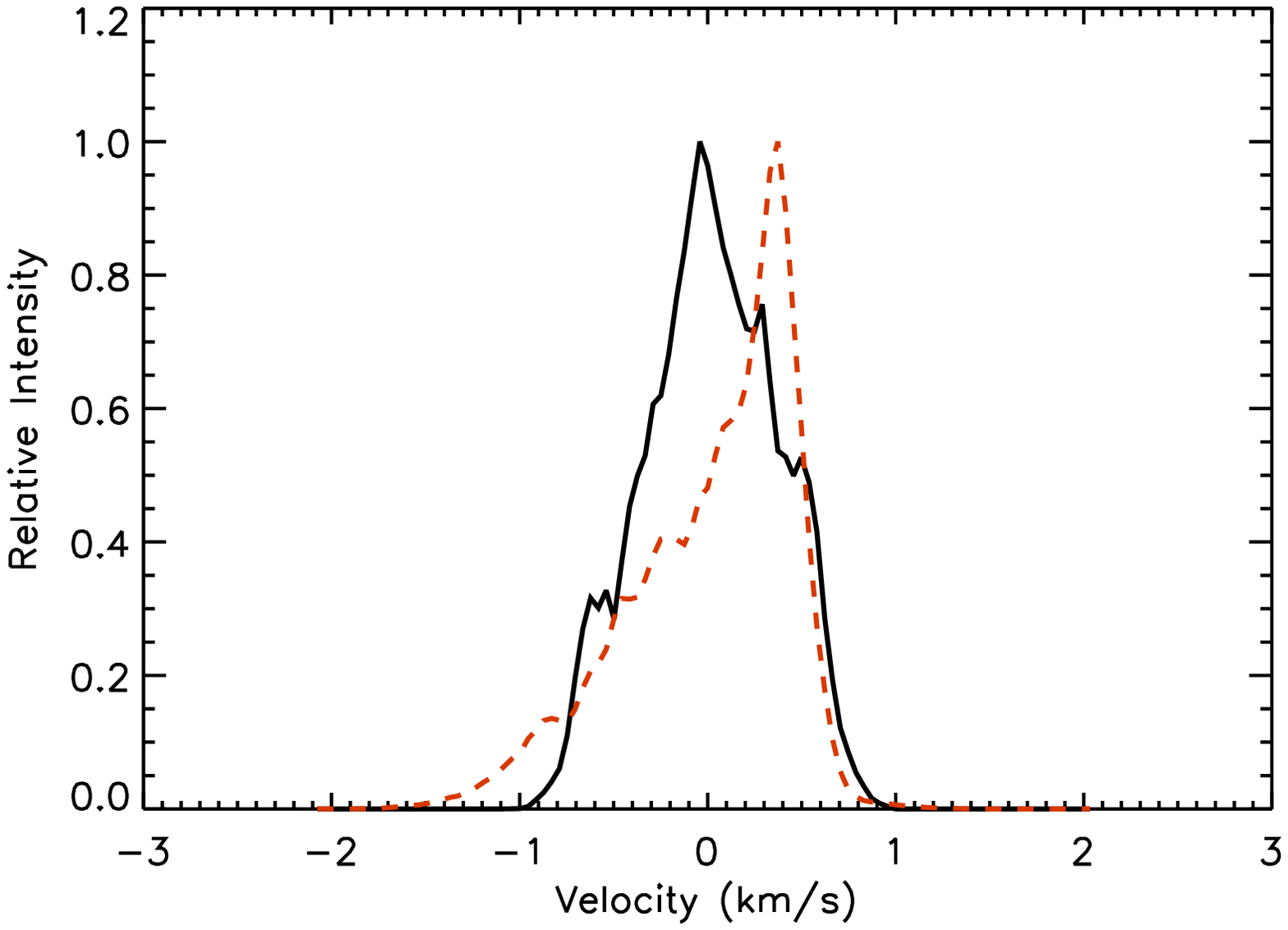}
}
\caption{\label{example2} As in Figure \ref{example}, except that the simulation has been smoothed to mimic the resolution of the observational taken by Walsh et al. (we assumed a FWHM resolution of 50'' at 140pc). Both the column-density contours (left panel) and velocity spectra (right panel) are smoothed slightly from those in Figure \ref{example}.}
\label{contoursoft}
\end{figure*}

\section{Methods of Analysis}
\label{method}

\citet{WalMyeBur2004} observed various low-mass star-forming molecular clouds using Doppler shift spectroscopy. By looking at specific molecular lines (tracers) in the spectra collected, they gathered information on the relative velocities of different components in the clouds. They chose the tracer N$_{2}$H$^{+}$ ($1-0$) to characterise the high-density gas within dense cores, and $^{13}$CO ($1-0$) and C$^{18}$O ($1-0$) to map the low-density gaseous envelopes surrounding the high-density cores. The N$_{2}$H$^{+}$ line is expected to trace molecular gas with densities $\sim 2\times 10^5$ cm$^{-3}$ \citep{Ungerechtsetal1997}.  CO lines trace gas with lower densities; we assume $\sim 10^3$ cm$^{-3}$ for $^{13}$CO \citep{Ungerechtsetal1997} and $\sim 10^4$ cm$^{-3}$ for C$^{18}$O \citep*{Tachiharaetal2000}.  To establish whether the core gases and envelope gases were in close proximity to one another along the line of sight their morphologies were compared. For close proximity to be assumed, they required ``that the peak of the CO emission lie within the 50\% contour of N$_{2}$H$^{+}$ and that the 50\% contour of the CO species encloses the 50\% N$_{2}$H$^{+}$ contour'' \citep{WalMyeBur2004}. The line-centre velocity difference between the high-density core and the low-density envelope indicates whether they are co-moving (small difference), or if the affiliation is short lived (large difference). They considered a large velocity difference to be similar to or greater than the CO line width, while a small difference should be less than the intrinsic N$_{2}$H$^{+}$ line width.

In order to make comparisons between these observations and the smoothed particle hydrodynamics (SPH) simulation of \citet{BatBonBro2003}, we analysed the simulations using a method that mimicked the observational analysis. We analysed 5 snapshots from the simulation taken at times of $t=1.0, 1.1, 1.2, 1.3$, and $1.4$ initial cloud free-fall times ($t_{\rm ff}=1.90\times 10^5$ yr).  The first snapshot is just before the first protostar forms, the last is at the end of the simulation when the cluster had formed 50 stars and brown dwarfs.  First, we separately selected the gas from the snapshots of the simulation that had densities between 1/3 and 3 times those of the typical tracer densities mentioned above (i.e., $2\times 10^5$, $10^3$ cm$^{-3}$, and $10^4$ cm$^{-3}$). For most of this paper, we assume there is little difference between $^{13}$CO and C$^{18}$O in the sense that both trace low-density gas and, therefore, we perform much of our analysis using only a single low-density tracer with a typical density of $10^3$ cm$^{-3}$.  However, for the final overall analysis we do consider two low-density tracers, associating $^{13}$CO with gas at $\sim 10^3$ cm$^{-3}$ and C$^{18}$O with gas at $\sim 10^4$ cm$^{-3}$.  We note that, in reality, the density-dependence of the tracer molecules will not be as sharp or distinct as the intervals we use to analyse the numerical simulations and may extend over a wider or narrower range of densities.  However, without actually including chemistry in the simulations it is difficult to make the analysis more accurate.  Furthermore, since the overall goal of this paper is to determine whether or not there is a velocity difference between the high-density cores and their low-density envelopes, using idealised densities and distinct cuts should make us {\it more likely} to detect the difference than is the case with observations in which signals from low- and high-density regions are somewhat blended.  Some examples of column-density plots of the density-screened clouds can be seen alongside full column-density maps in Fig.~\ref{pics} (note that the high-density gas and low-density gas are plotted together in each of the centre and right-hand panels, rather than plotting them separately, since this allows a visual comparison of where they are situated relative to each other).  

\begin{figure*}
\centering
\subfigure 
{
    \includegraphics[width=5.2cm]{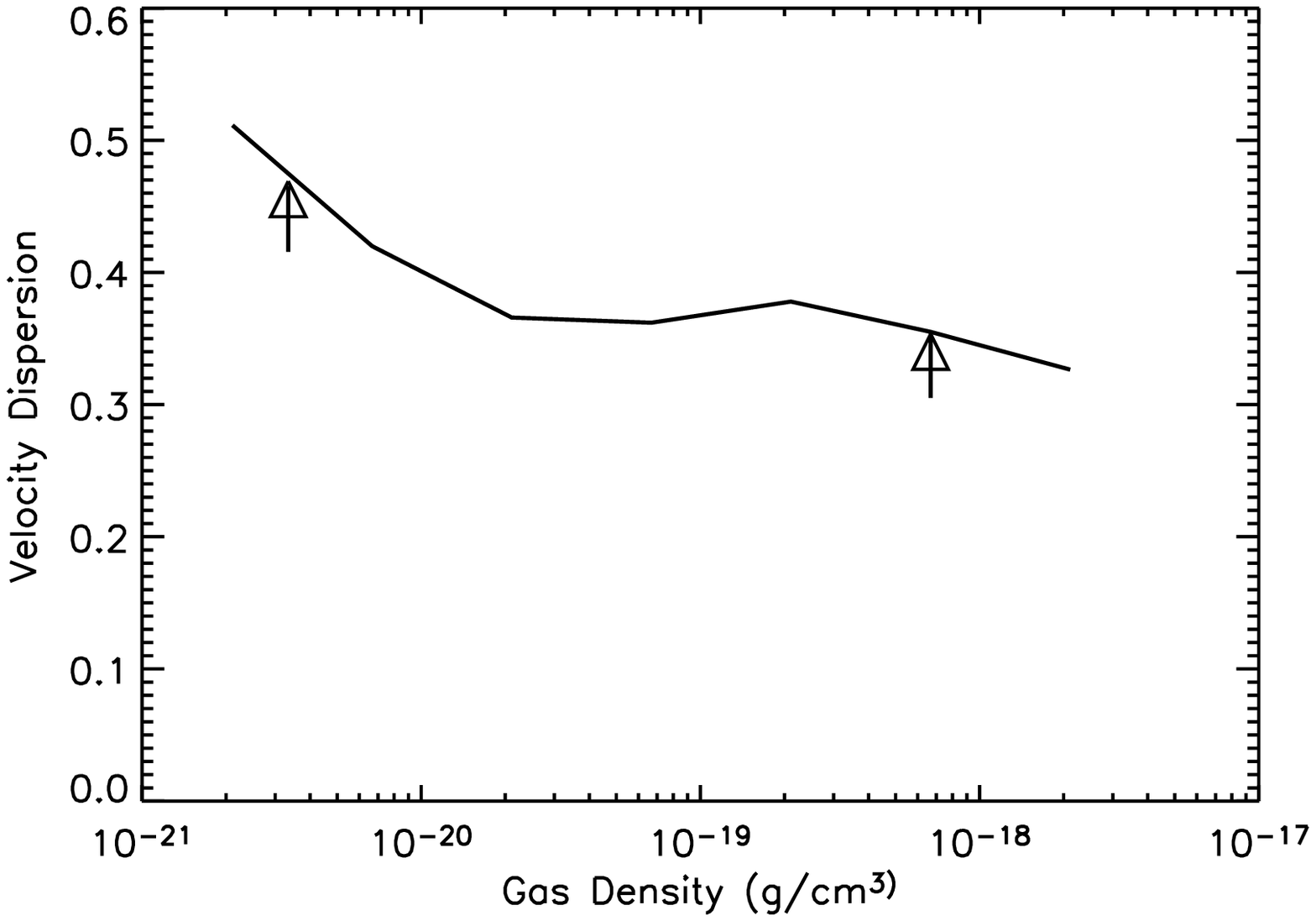}
}
\subfigure 
{
    \includegraphics[width=5.5cm]{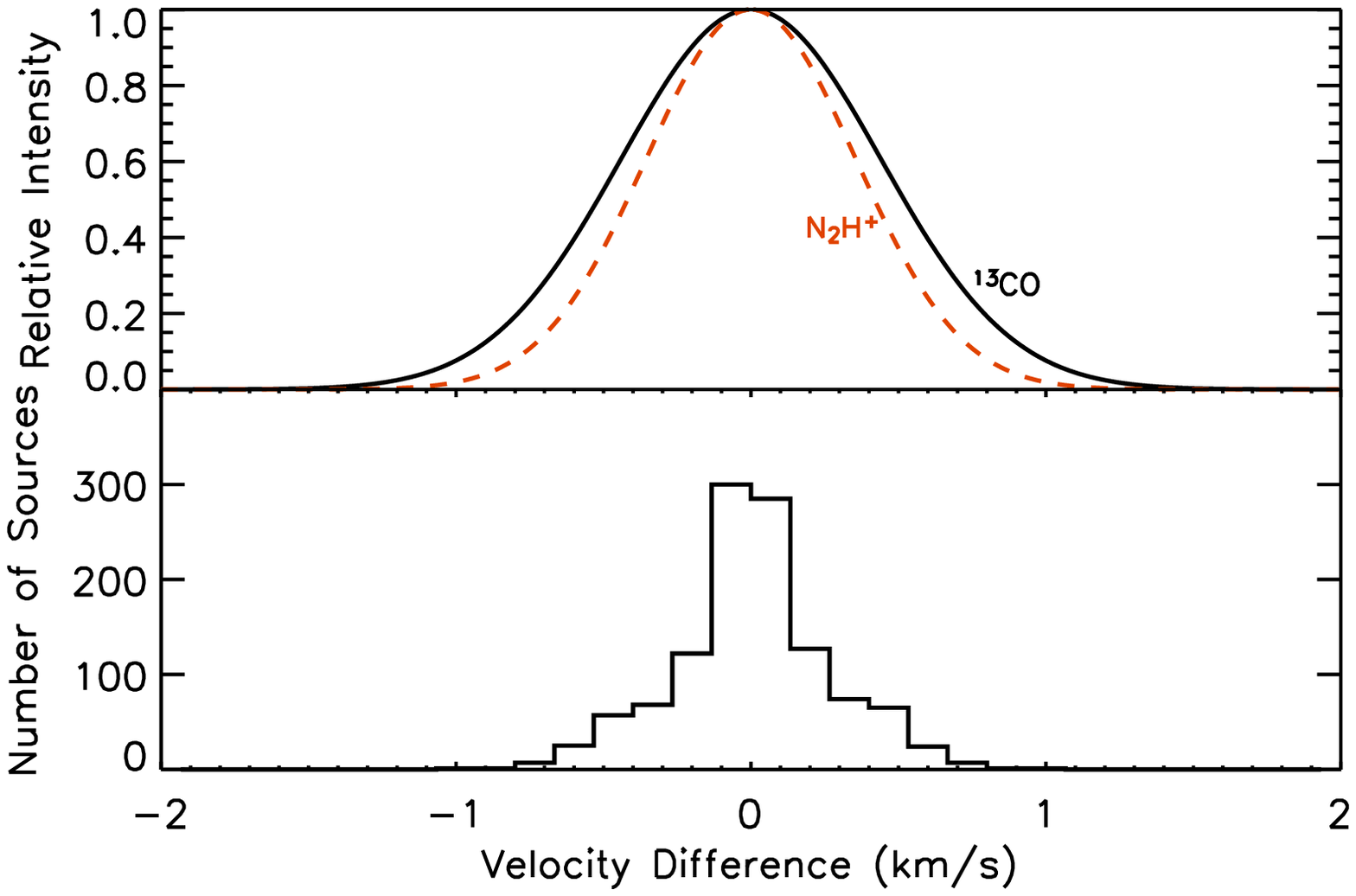}
}
\subfigure 
{
    \includegraphics[width=5.5cm]{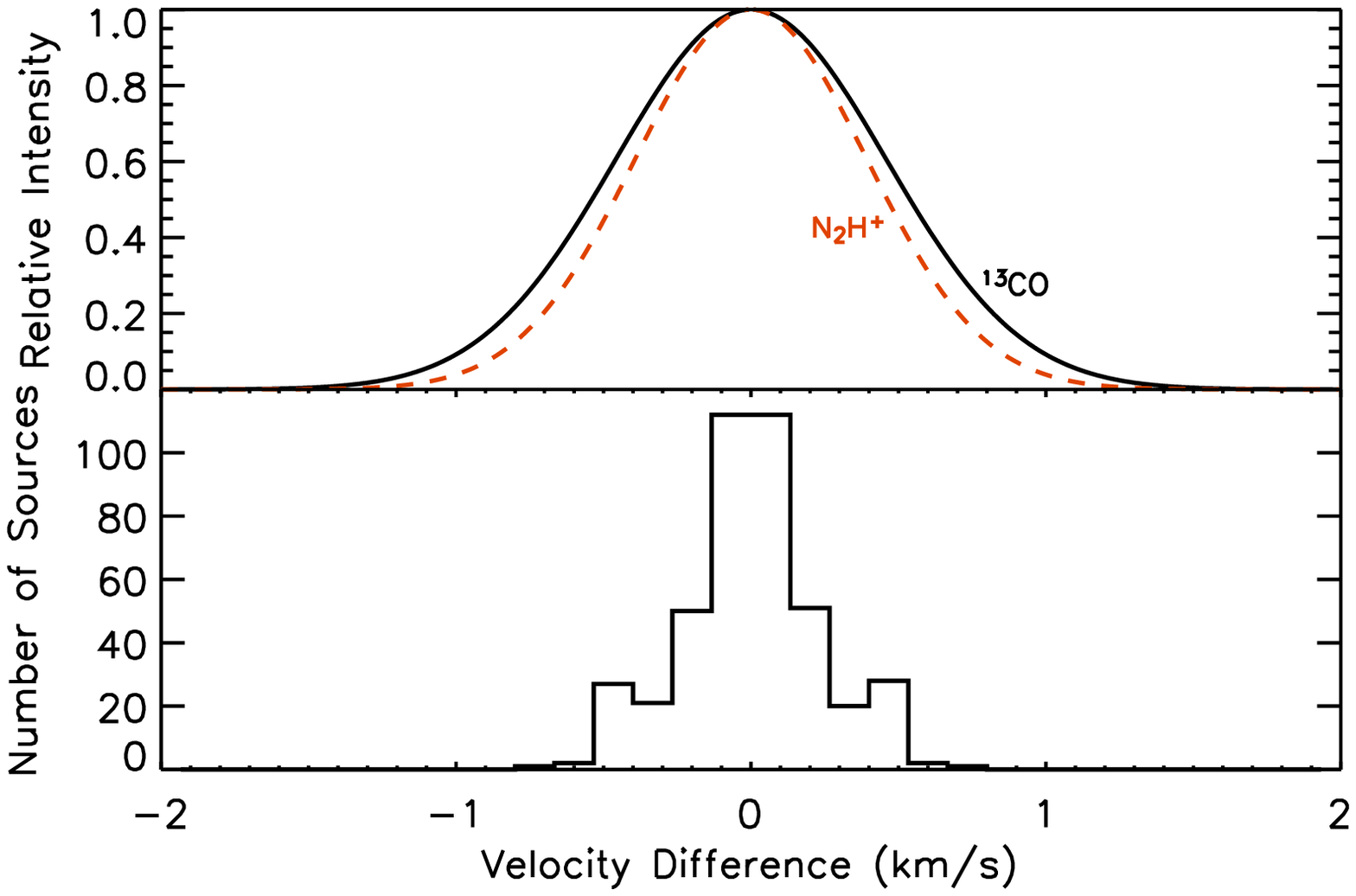}
}

\vspace{-0.2cm}

\subfigure 
{
    \includegraphics[width=5.2cm]{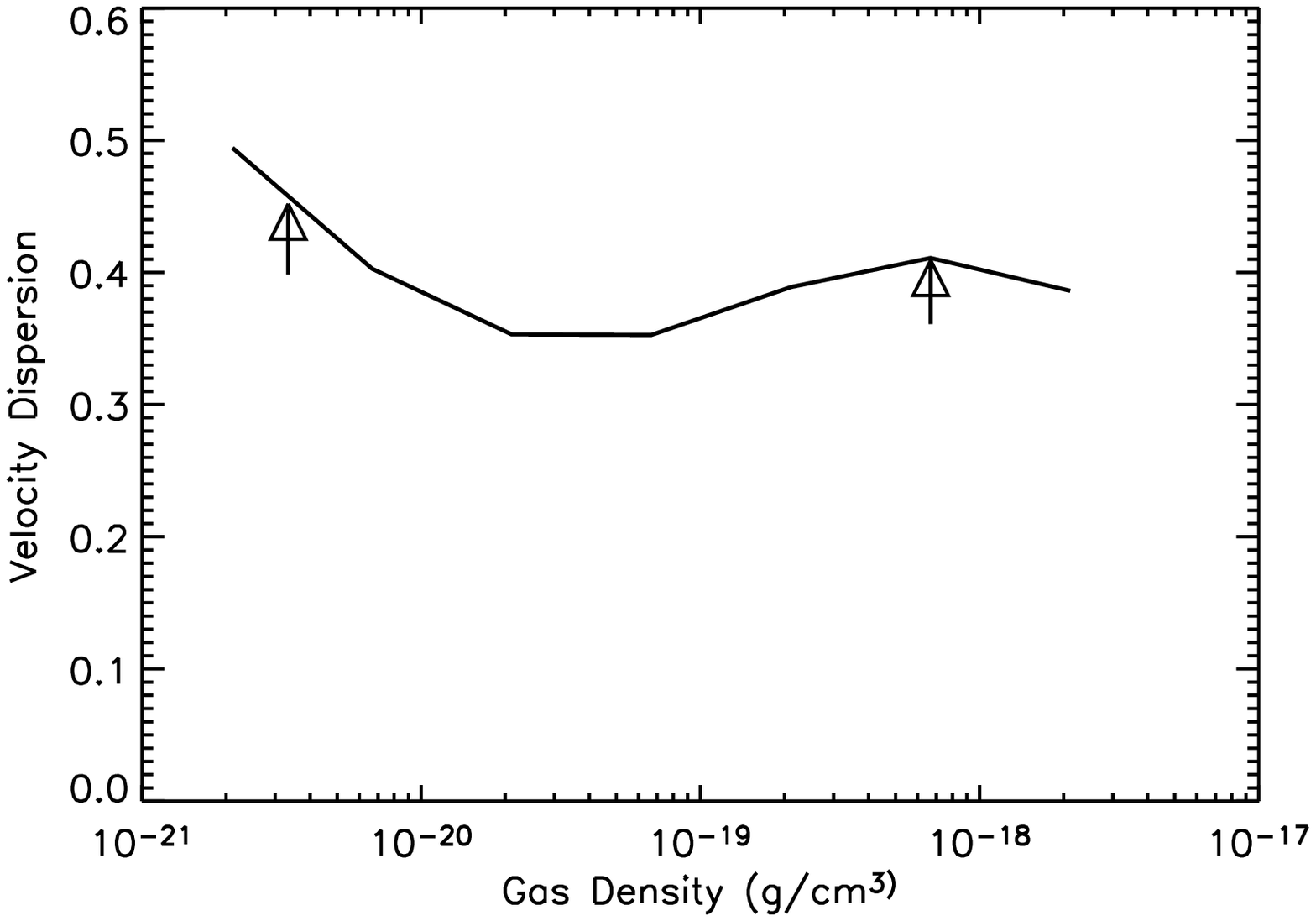}
}
\subfigure 
{
    \includegraphics[width=5.5cm]{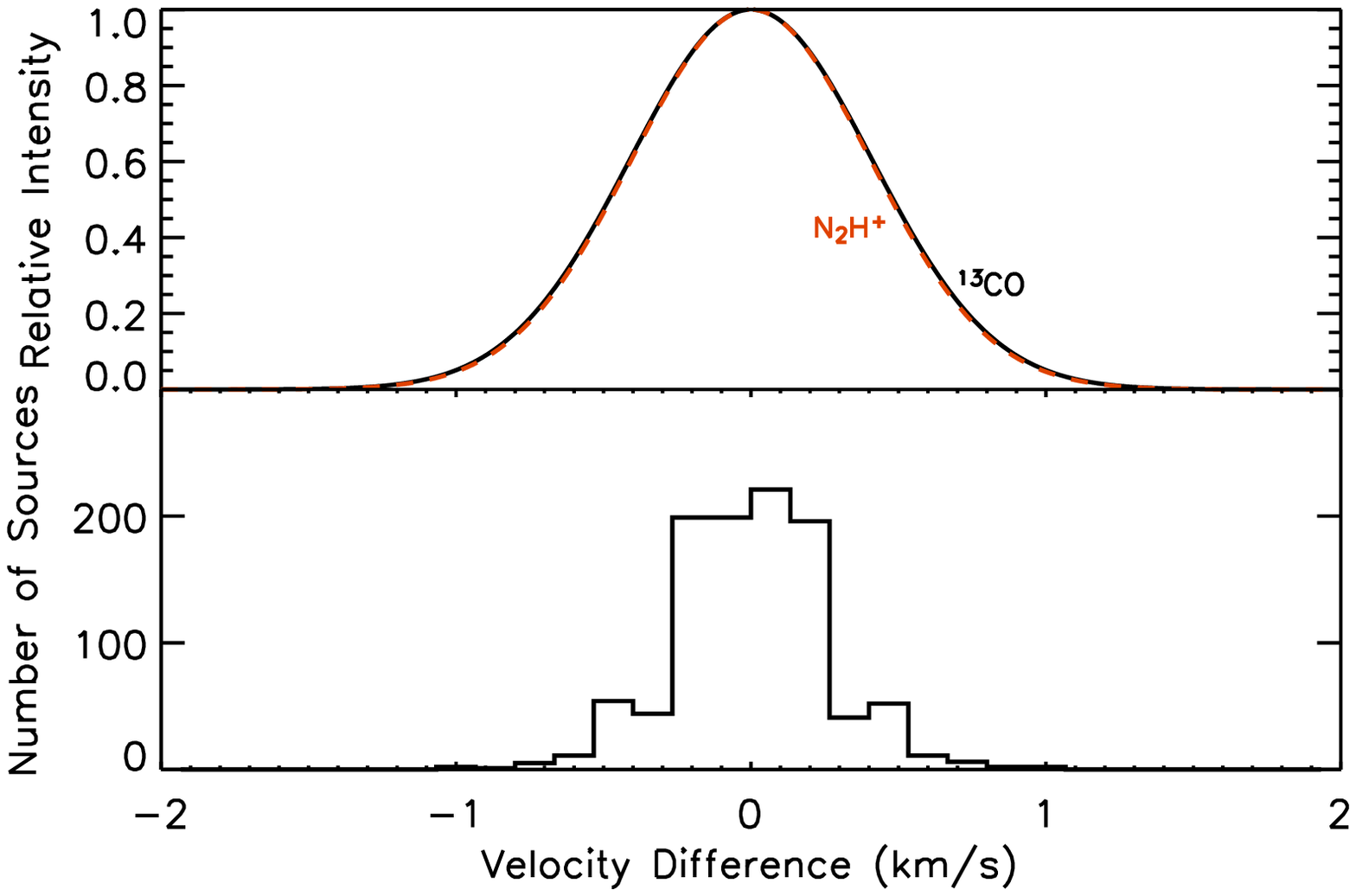}
}
\subfigure 
{
    \includegraphics[width=5.5cm]{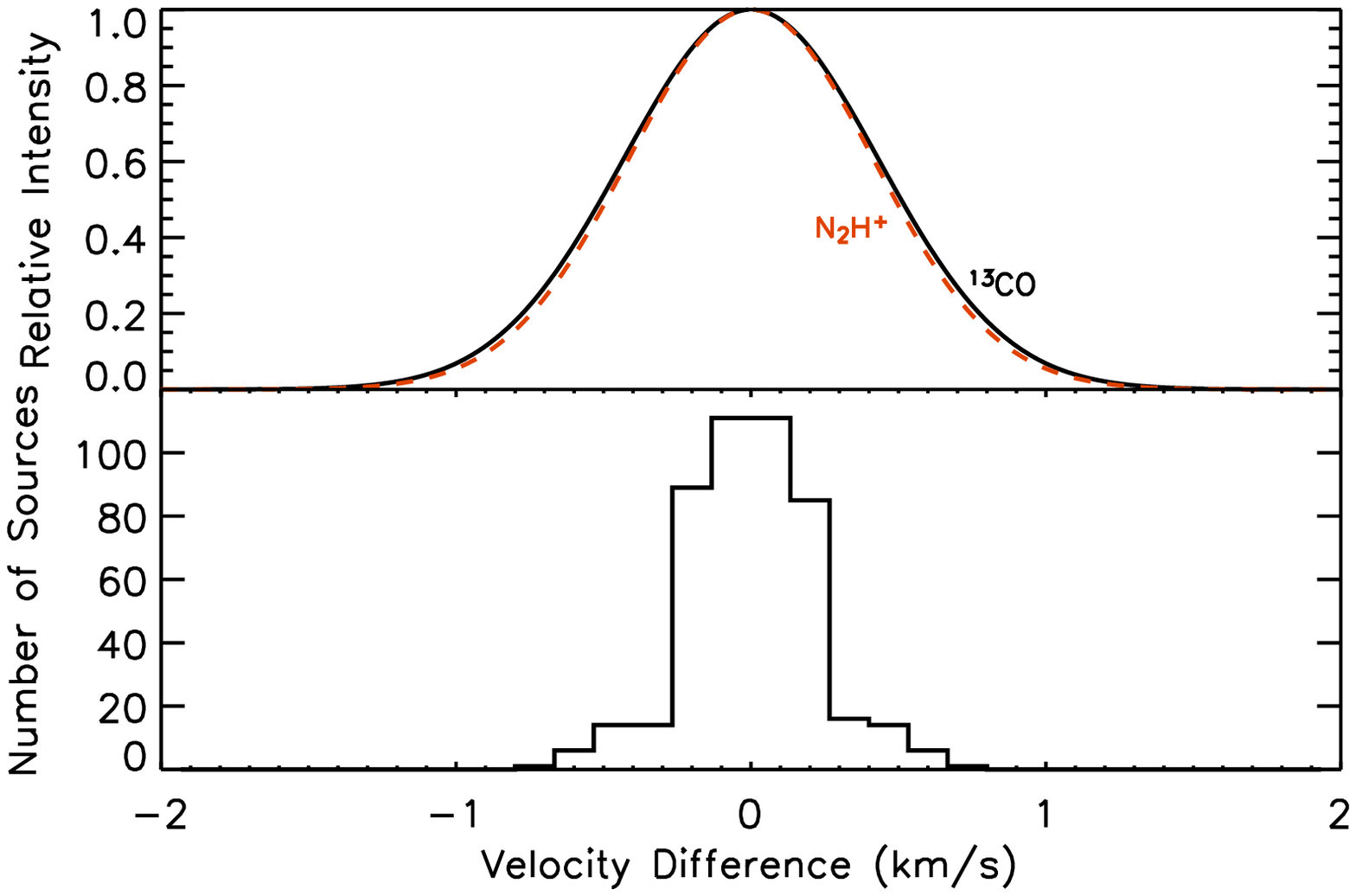}
}

\vspace{-0.2cm}

\subfigure 
{
    \includegraphics[width=5.2cm]{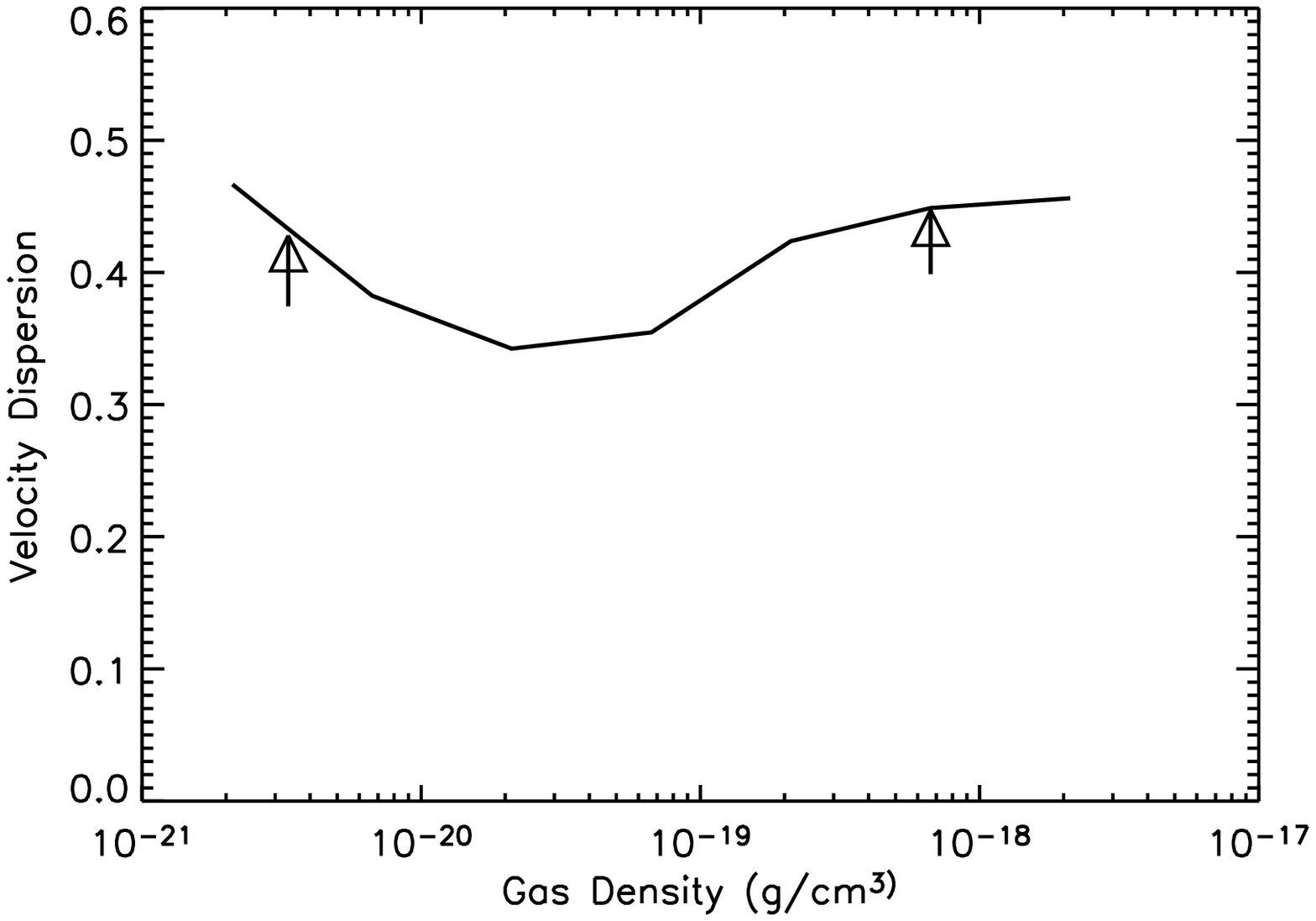}
}
\subfigure 
{
    \includegraphics[width=5.5cm]{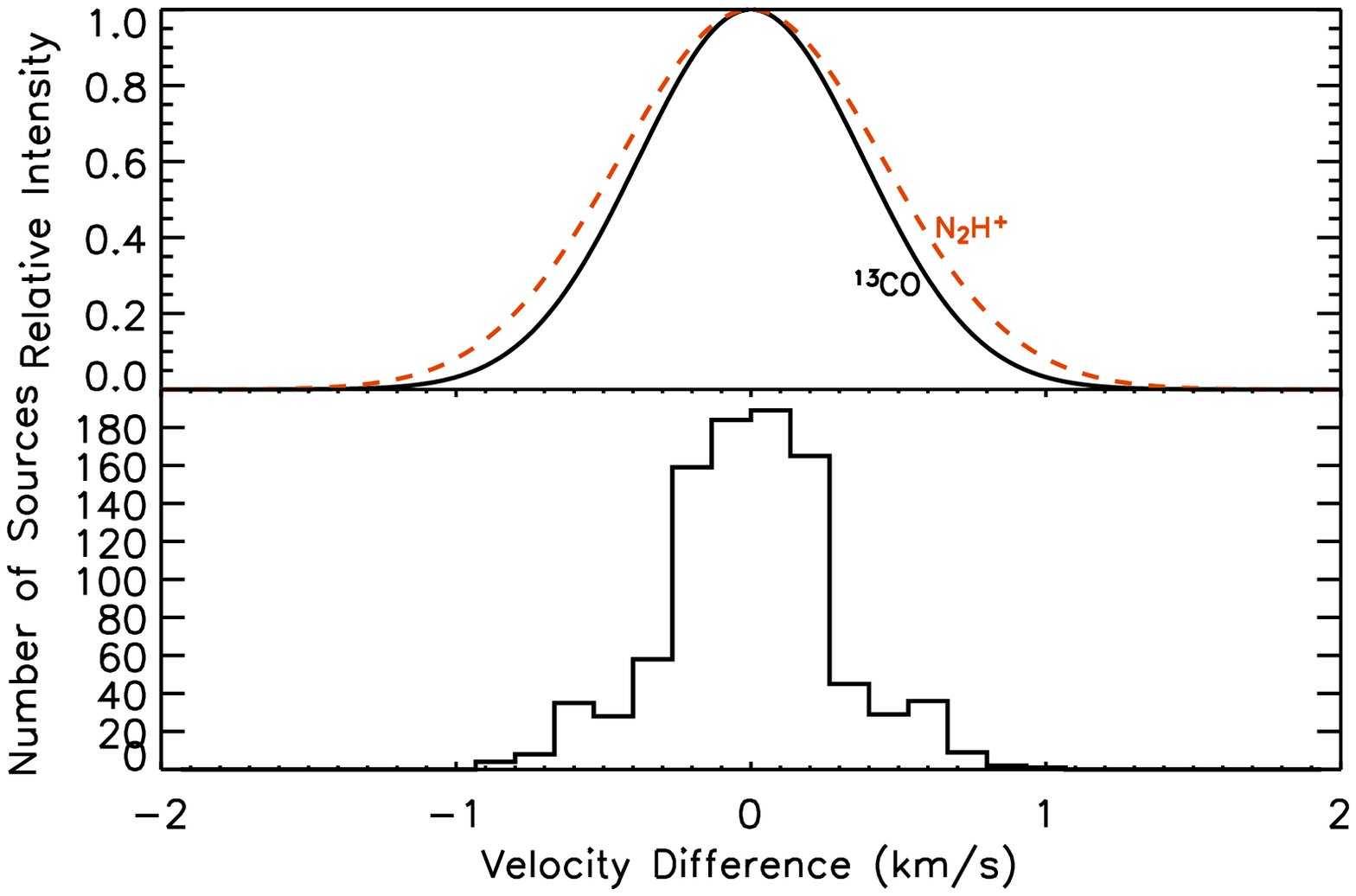}
}
\subfigure 
{
    \includegraphics[width=5.5cm]{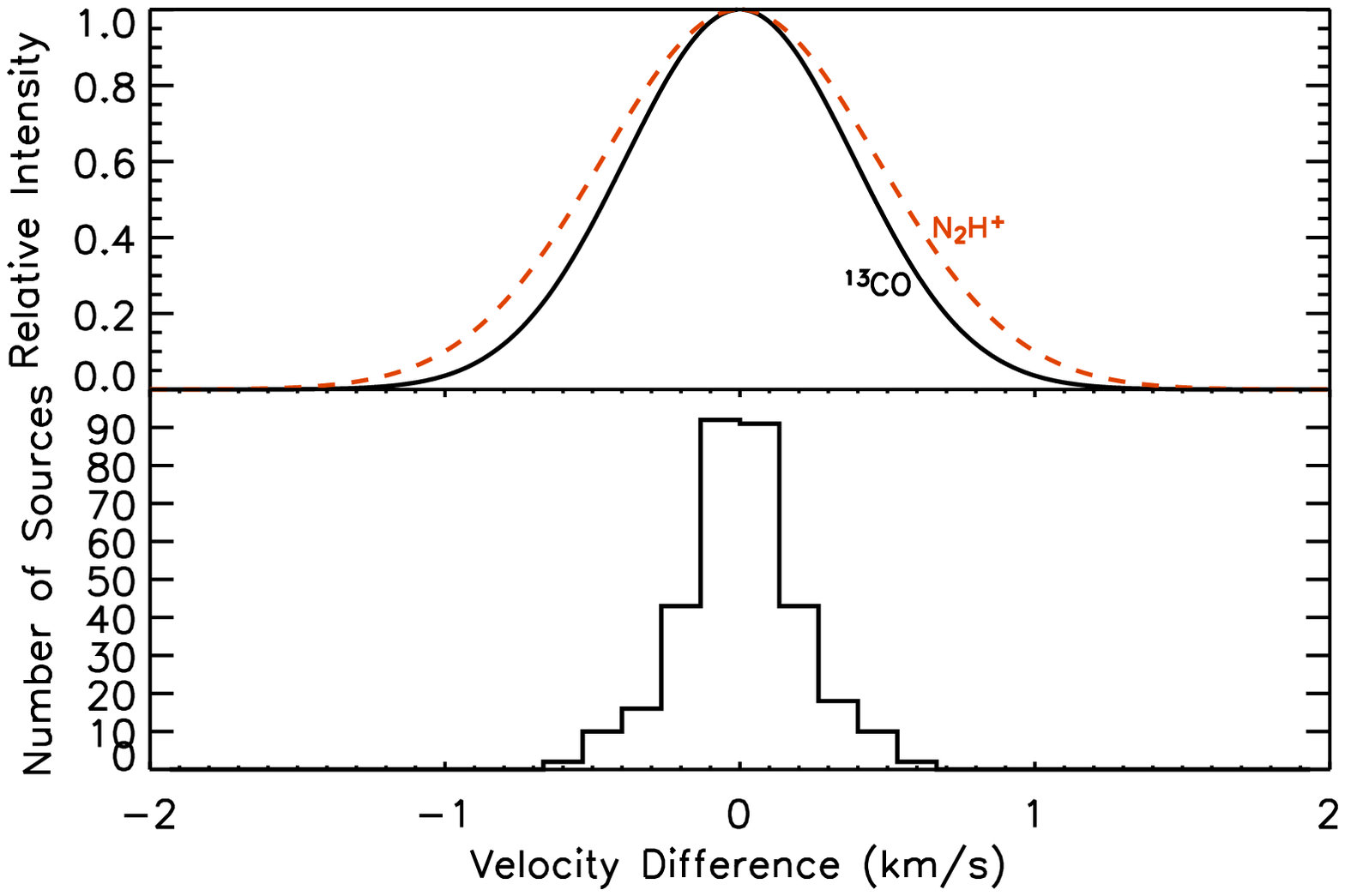}
}

\vspace{-0.2cm}
\subfigure 
{
    \includegraphics[width=5.2cm]{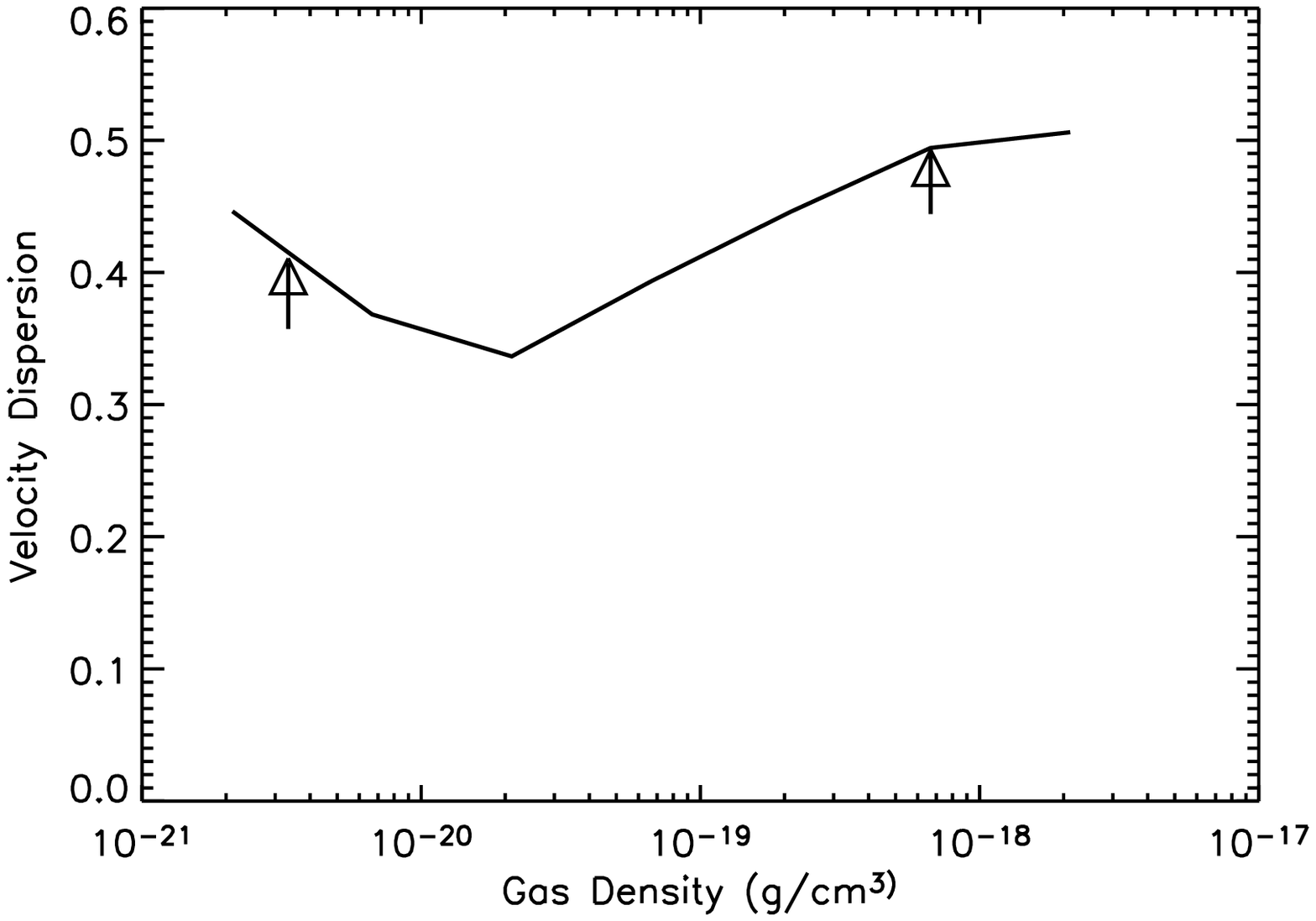}
}
\subfigure 
{
    \includegraphics[width=5.5cm]{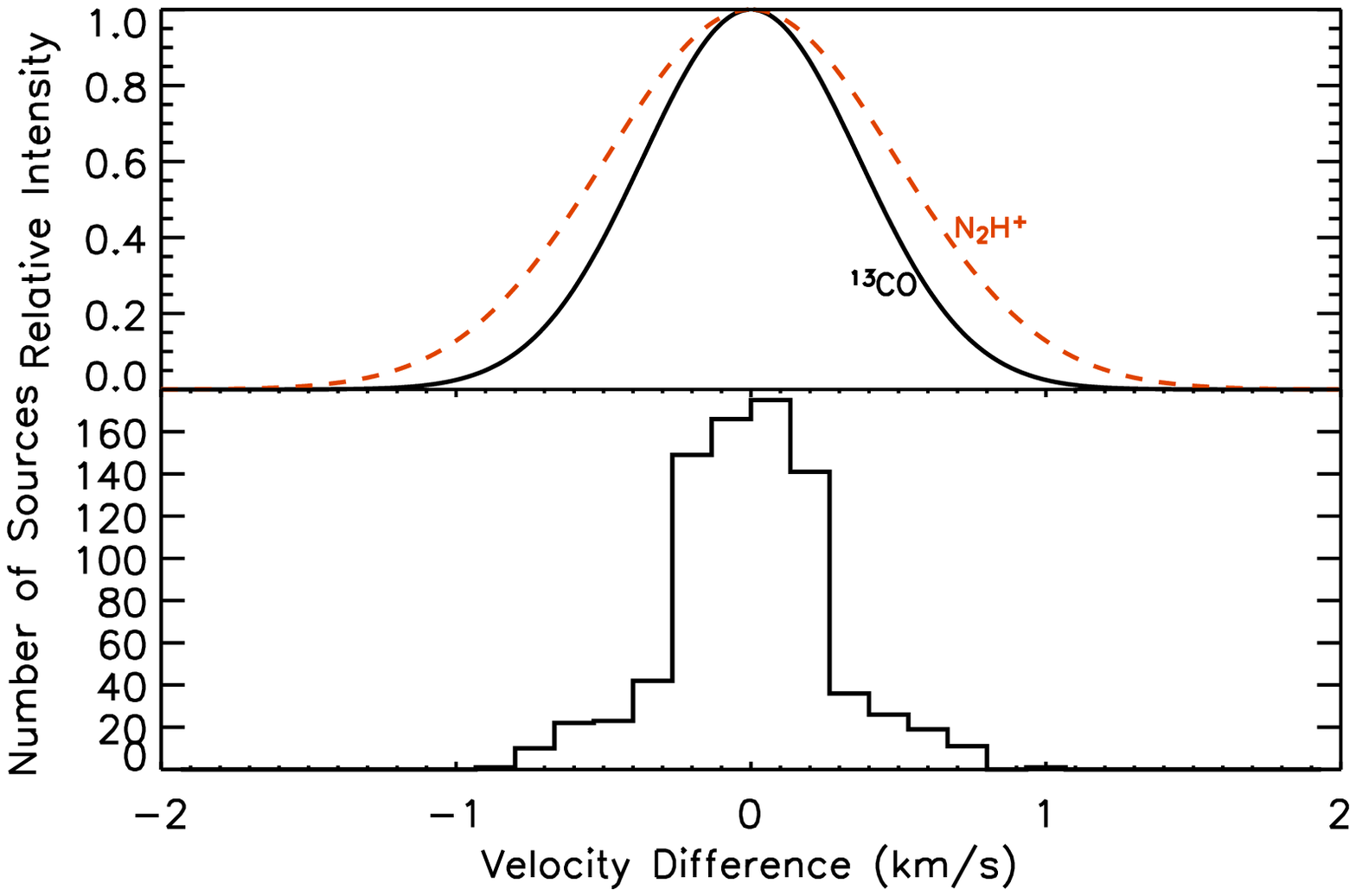}
}
\subfigure 
{
    \includegraphics[width=5.5cm]{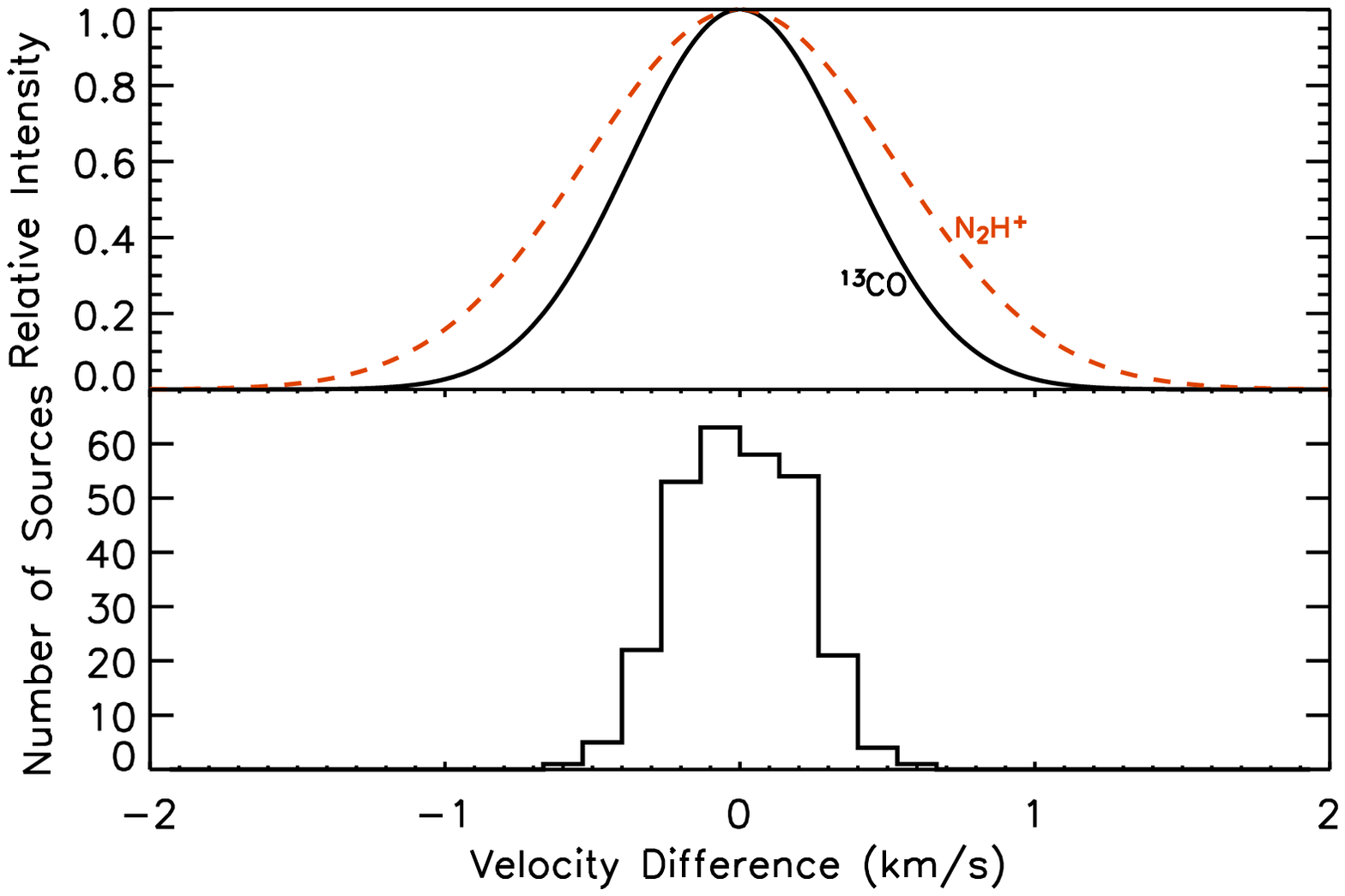}
}

\vspace{-0.2cm}

\subfigure 
{
    \includegraphics[width=5.2cm]{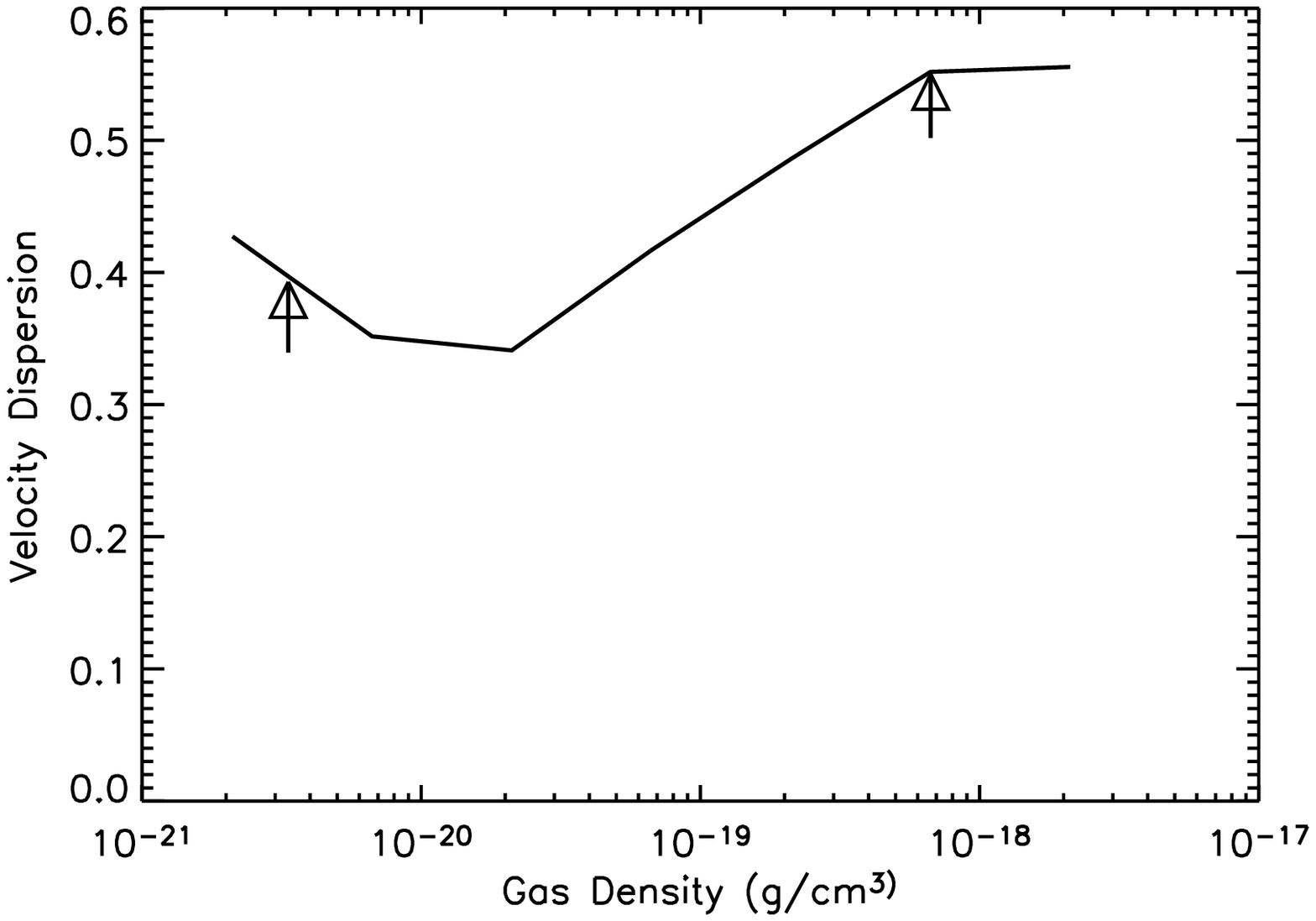}
}
\subfigure 
{
    \includegraphics[width=5.5cm]{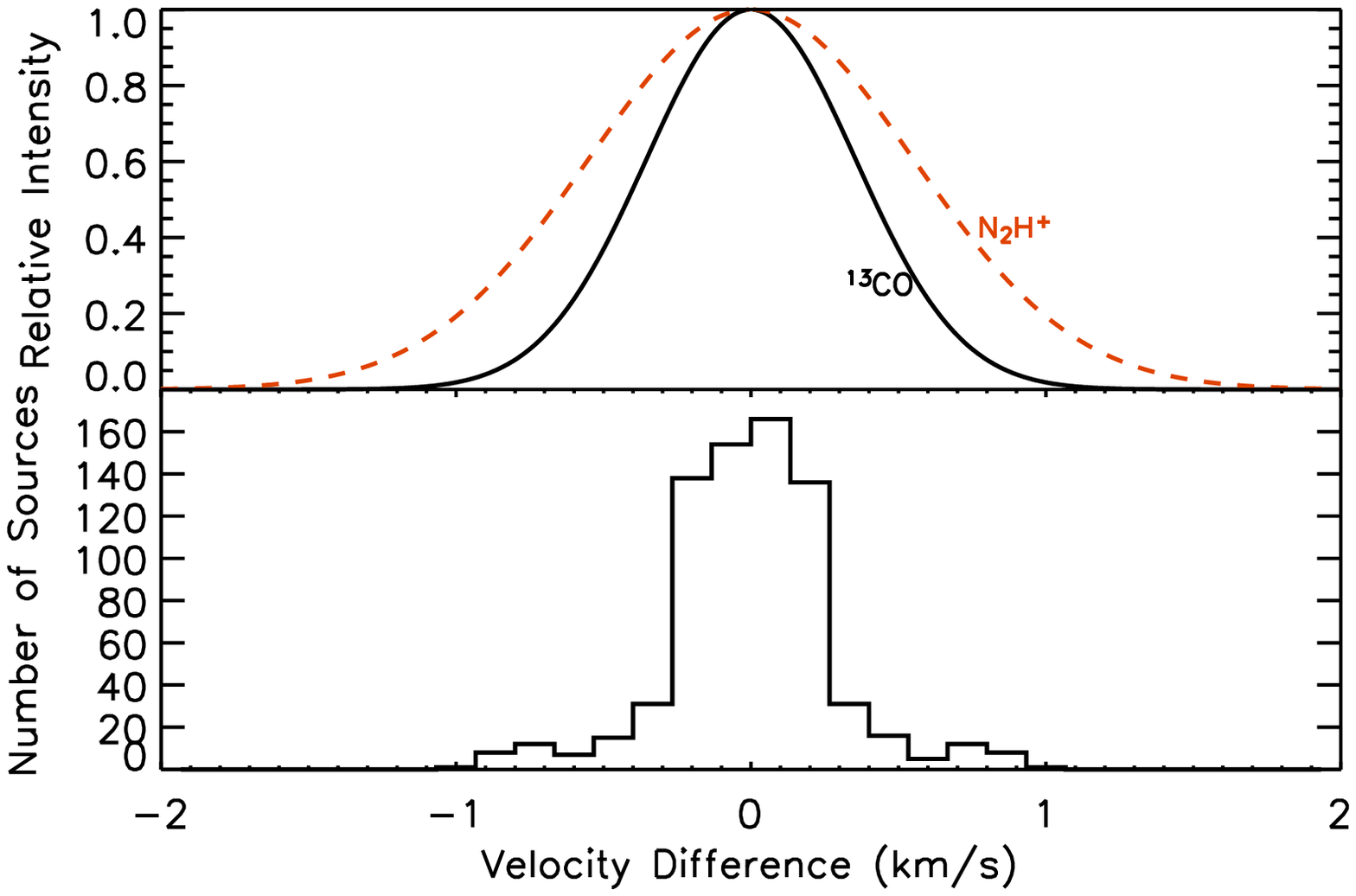}
}
\subfigure 
{
    \includegraphics[width=5.5cm]{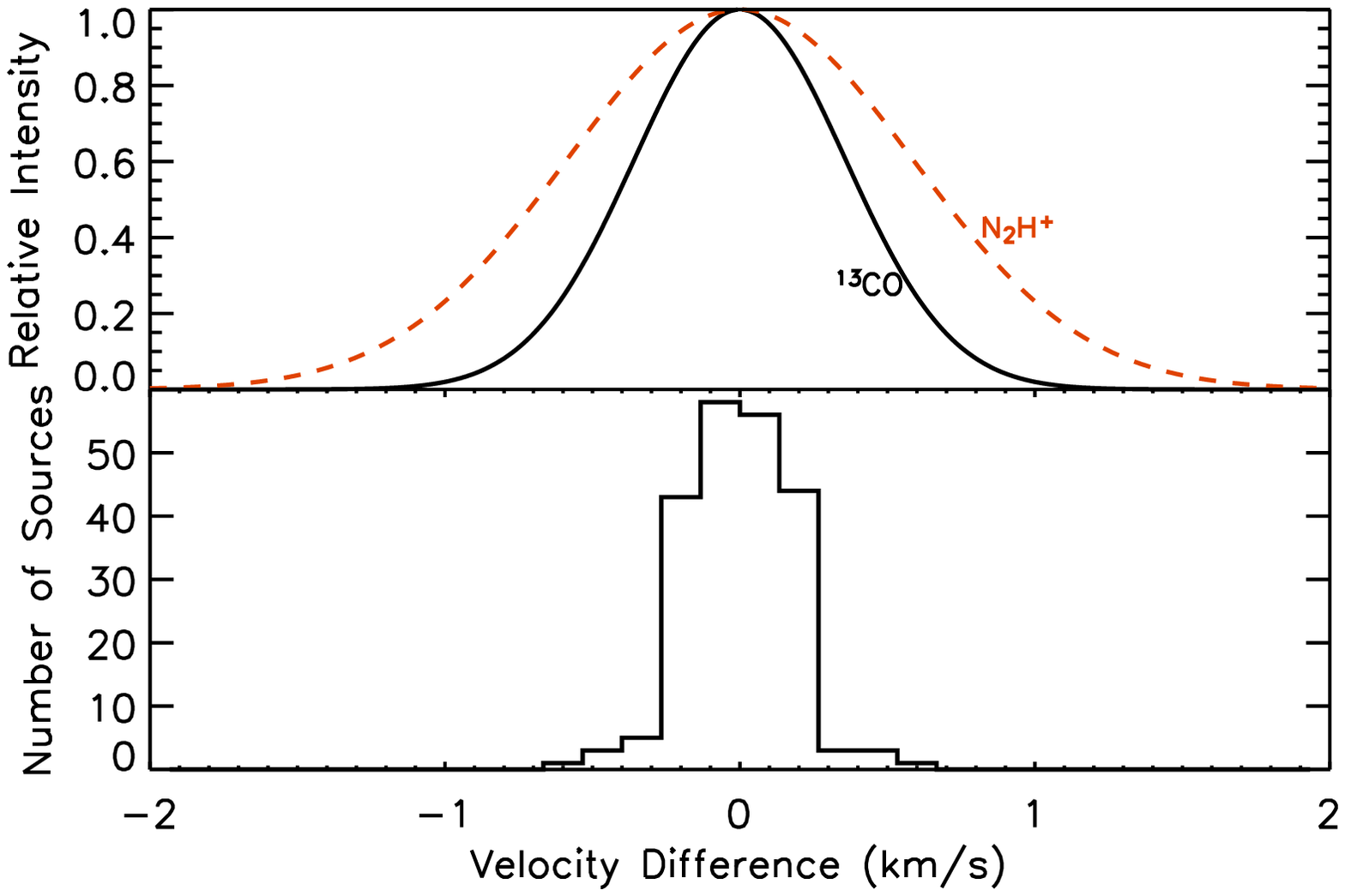}
}

\caption{\label{densplots} The left panels give the velocity dispersion as a function of gas density at five different times during the simulation ($t=1.0, 1.1, 1.2, 1.3,$ and 1.4 $t_{\rm ff}$, top to bottom). In the centre panels, the upper plot at each time displays the velocity dispersion using representative Gaussians at the densities denoted by the arrows in the left hand plots (i.e., high density of $2 \times 10^5$~cm$^{-3}$, red dashed line; low density of $1 \times 10^3$~cm$^{-3}$, black solid line).  The lower panel at each time displays a histogram of the line-centre velocity differences between the high-density core and low-density envelope gas.  These figures present our results in the same manner as \citet{WalMyeBur2004} (Figure 2 of their paper).  The right panels give the same information as the centre panels, but were produced using the smoothed data.  In all cases, the histograms are much narrower (with dispersions ranging from 0.25 to 0.27 km~s$^{-1}$) than the high- or low-density gas velocity dispersions (ranging from 0.35 to 0.55 km~s$^{-1}$), and in all cases, smoothing of the data to match the observational resolution decreases the dispersion of the line-centre velocity differences (which then range from 0.18 to 0.24 km~s$^{-1}$).}
\end{figure*}

For a particular line-of-sight towards the cloud, we constructed column-density maps ($100\times 100$ pixels over spatial scales of $0.8\times 0.8$ pc) for each gas density.  Applying the same proximity relation (mentioned above) as \citet{WalMyeBur2004} to these column-density maps yielded cores and envelopes meeting the same criteria as in the observations. This relation requires that intensity peaks be identified for each of the tracers. This was achieved utilising a simple neighbouring pixel comparison, scanning the entire field and identifying points with intensity values higher than the 4 nearest neighbouring points. Contours about the high-density gas peaks, at 50\% of the peak's value, were then identified. Coordinates within these contours were then checked for any low-density gas peaks. If found, a contour at 50\% the value of the maximal low-density gas column-density peak was then identified. In the analysis of the simulation, most high-density contours contained a low-density peak and passed the first test -- only 22 percent of the high-density cores failed.  The second and final requirement was that the low-density contour entirely encompass the high-density contour. Should the final criteria fail to be met by the 50\% contour of the maximal low-density gas peak, the next highest low-density gas peak was used to produce a contour, and so on, until all the peaks were exhausted. The vast majority of cores passed this second test -- only 3 percent failed. If the final criteria was fulfilled then the high-density core (N$_{2}$H$^{+}$ equivalent) and low-density envelope ($^{13}$CO equivalent) were considered to be associated with one another (as in the observational analysis).  When we considered both $^{13}$CO and C$^{18}$O envelopes, these low-density regions both had to pass the above tests but there was no additional criteria applied between the two low-density tracers.

The radial velocities of SPH gas particles contributing to each pixel inside the 50\% high-density contour were then added together, weighted by their contribution to the line-of-sight column densities, to produce two velocity spectra, one each for the high- and low-density gas. This gave the equivalent of Doppler spectroscopy maps of the high- and low-density gas. The `line-centre' velocities of the high-density core and low-density envelopes were then calculated as the mean velocities of these `spectra'.  

An example of an extracted core, its corresponding envelope, and their velocity spectra are shown in Fig.~\ref{example}. Typically, viewing a snapshot from a particular angle resulted in 7 or 8 cores being identified. The average core and envelope masses were $9$ and $5$ M$_\odot$ and their average diameters were 0.16 and 0.6 pc, respectively. These are similar to the sizes of the cores and envelopes studied by Walsh et al., and the masses of the cores are similar to those observed in low-mass star-forming regions.  Although it might have been expected that the typical envelope mass would be larger than the typical core mass, we caution against over-interpreting these numbers -- the values obtained are entirely dependent on the density ranges that we assume when taking the low-density and high-density cuts.  Furthermore, recall that the simulation being analysed here only contains 50 M$_\odot$ in total.  Later in the simulation $\approx 10$ percent of this mass is already in the form of stars and brown dwarfs, while much of the remaining gas falls outside of the two density cuts.  Thus, with a typical core mass of 9 M$_\odot$, the mass defined as being in the envelope is never likely to be substantially larger than the typical core mass.

We view each of the 5 snapshots from $\approx 80$ different angles. From each different view point both the shapes of the cores/envelopes and the motions of the gas along the line-of-sight change.  This not only increases our sample size, but also ensures that we will detect high relative velocities (if they exist) even if the relative velocities are apparently small from some particular lines-of-sight.   So by viewing the clouds from multiple directions we obtain many distinct sets of results.

At the suggestion of the referee, we also examined the effect of degrading the resolution of the column-density maps to match observations.  The beam widths of the observations used by \citet{WalMyeBur2004} are quoted as being 35" or 50" FWHM depending on the observations.  In the SPH simulations, particles are inherently smoothed over their hydrodynamic smoothing (resolution) length using a spline function that approximates a Gaussian.  Thus, smoothing the data to mimic the observations was easily accomplished simply by taking the greater of the hydrodynamical smoothing length from the simulation and half (since 50" is the diameter of the beam rather than the radius) the distance that subtends an angle of 50" at 140 pc (i.e., 0.017 pc) when performing the analysis.  An example of the effect of the smoothing is given in Fig.~\ref{example2}. We expected that any smoothing of the data (mixing in gas from nearby spatial locations) was only likely to {\it decrease} the magnitude of the measured relative velocities between the cores and envelopes.  This was true in every case (see Figure \ref{densplots} and the next section).

\begin{figure}
  \mbox{\includegraphics[width=8cm]{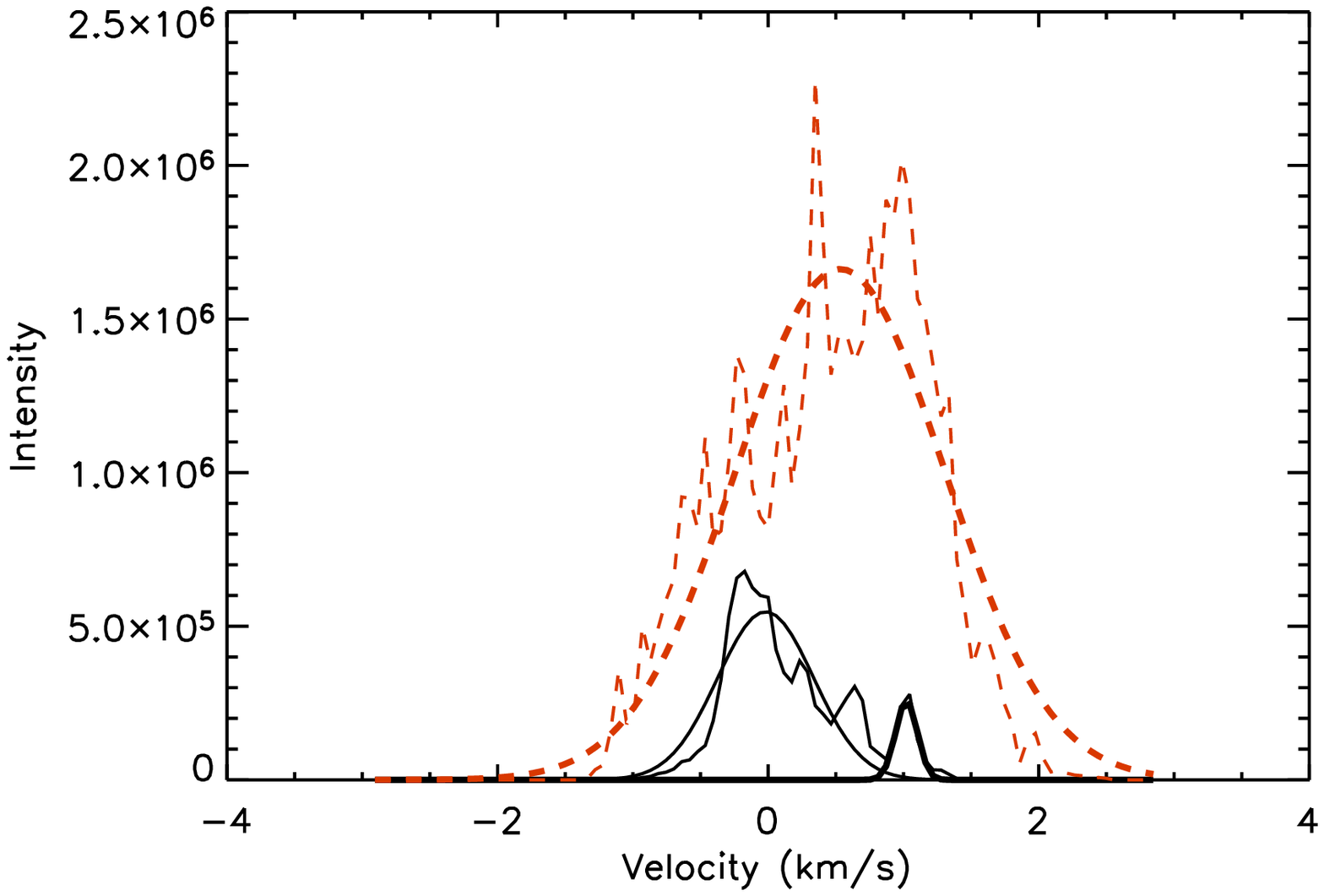}}
\caption{\label{gauss} An example of the high-density gas velocity `spectrum' with a Gaussian fit (red dashed lines) and the corresponding low-density envelope velocity `spectrum' with fit by two Gaussians (black solid lines).  Using Gaussians to fit the envelope velocity `spectrum' often does not result in a good fit.  Therefore, as discussed in the main text, we elected simply to calculate the mean and dispersion of each velocity spectrum rather than trying to extract a velocity line-centre and dispersion from Gaussian fits.}

\end{figure}

\begin{figure}
  \mbox{\includegraphics[width=8cm]{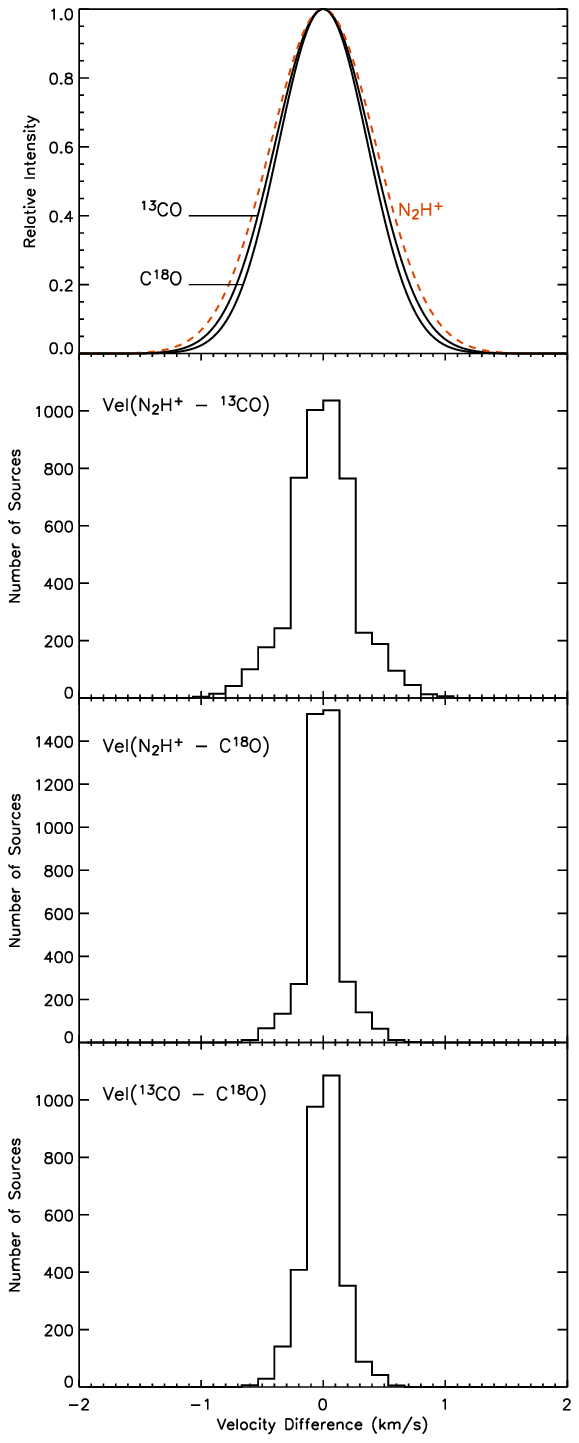}}
\caption{\label{hists} These plots give the overall results from the analysis of the simulation snapshots averaged over all viewing angles and all five times. Following Walsh et al.\ (Figure 2 of their paper) in the upper panel, we plot three Gaussian distributions whose standard deviations are given by the mean standard deviations of all of the high-density core (equivalent to N$_2$H$^+$) and low-density envelope (equivalent to $^{13}$CO and C$^{18}$O) velocity spectra.  The velocity dispersions are 0.43, 0.40, and 0.37 km~s$^{-1}$, respectively. In the lower panels, we plot histograms of the distribution of differences in mean (equivalent to line-centre) velocities between each pair of tracers (gas densities).  The dispersions of the histograms are 0.27, 0.16, and 0.16 km~s$^{-1}$, from top to bottom.}
\end{figure}

\citet{WalMyeBur2004} determined whether cores and envelopes had lasting a affiliation (i.e., were moving slowly relative to each other) by comparing the line-centre velocity distributions of the different tracer species. They determined line-centre velocities using Gaussian fits to the velocity spectra given by combining each individual spectrum from each pointing within the cores/envelopes. They state that low-density gas velocity distributions that could be fit with multiple Gaussian components were fitted with two or three Gaussians, and that the Gaussian with the line-centre velocity that was closest to the N$_{2}$H$^{+}$ line-centre velocity was selected as representing the associated envelope. We took two approaches to dealing with our synthetic spectra.  One was to use a method similar to Walsh et al., fitting multiple Gaussians.  So for each case we fitted a Gaussian to a low-density spectrum and if the peak of the residual was greater than 25\% of the original peak, we fitted a second Gaussian.  However, the spectra of our low-density gas envelopes were often not well represented by one or two Gaussians but displayed more structure (e.g. Fig.~{\ref{gauss}).  Furthermore, when Walsh et al.\ fitted more than one Gaussian to their low-density gas distributions, they always took the Gaussian whose line-centre velocity was closest to that of the high-density tracer as being the envelope that was associated with the high-density core.  Since their method of looking for large relative motions depends entirely on the distribution of relative line-centre velocities, this seemed to us to potentially bias their results in favour of small relative velocities.  Therefore, we decided that a more robust method was not to attempt to fit Gaussians to give a line-centre and a dispersion, but instead simply to calculate the mean (line-centre) velocity and the standard deviation (dispersion) of the spectrum about this mean.  In fact, when averaged over the five different snapshots and multiple viewing angles, both methods of analysis gave similar conclusions.  However, because we believe that the second method is more robust, we only discuss those results from this point on.

\begin{figure*}
\centering
\subfigure 
{
    \includegraphics[width=5.5cm]{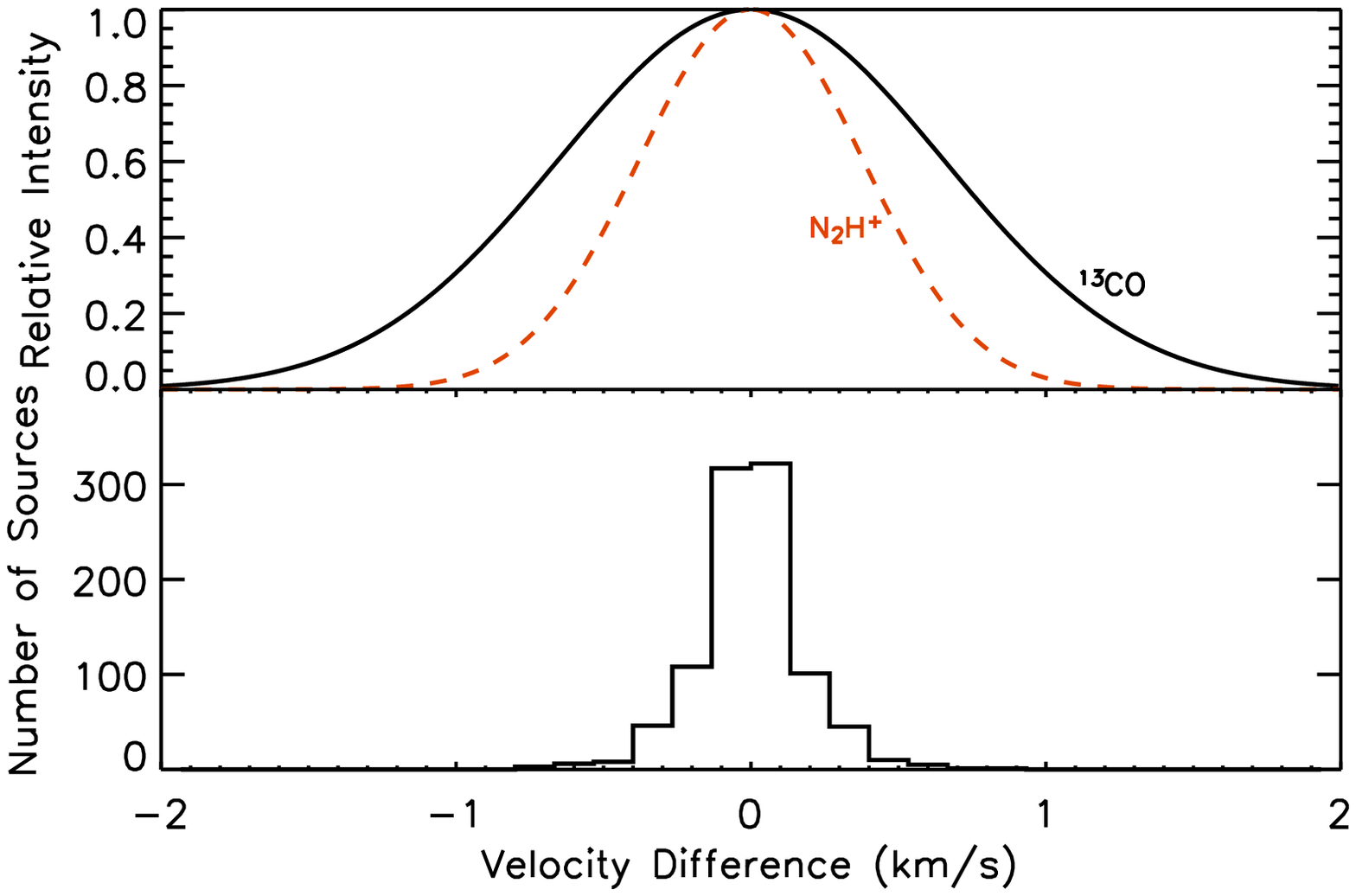}
}
\subfigure 
{
    \includegraphics[width=5.5cm]{B246.eps}
}
\subfigure 
{
    \includegraphics[width=5.5cm]{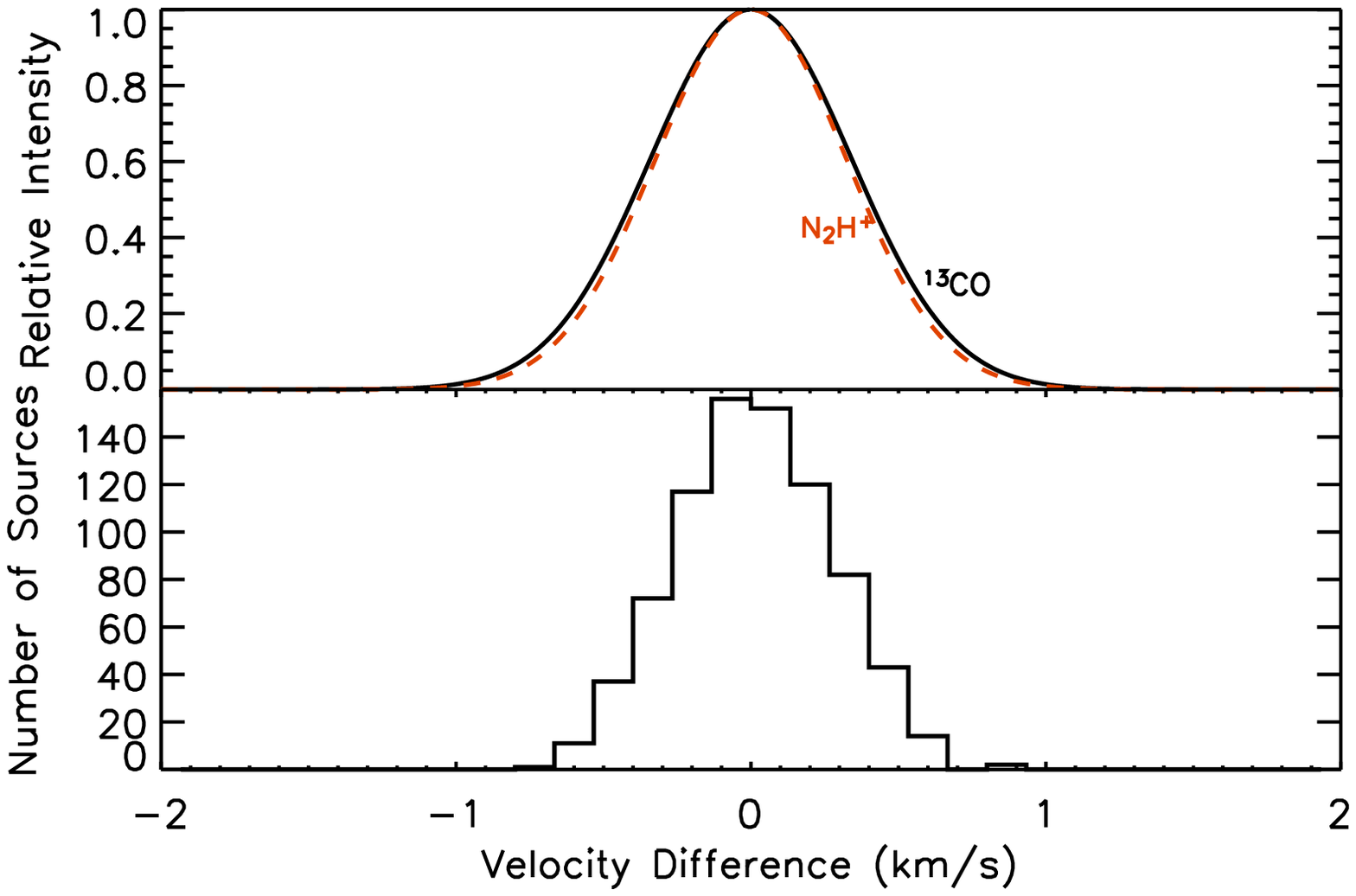}
}

\vspace{-0.2cm}

\subfigure 
{
    \includegraphics[width=5.5cm]{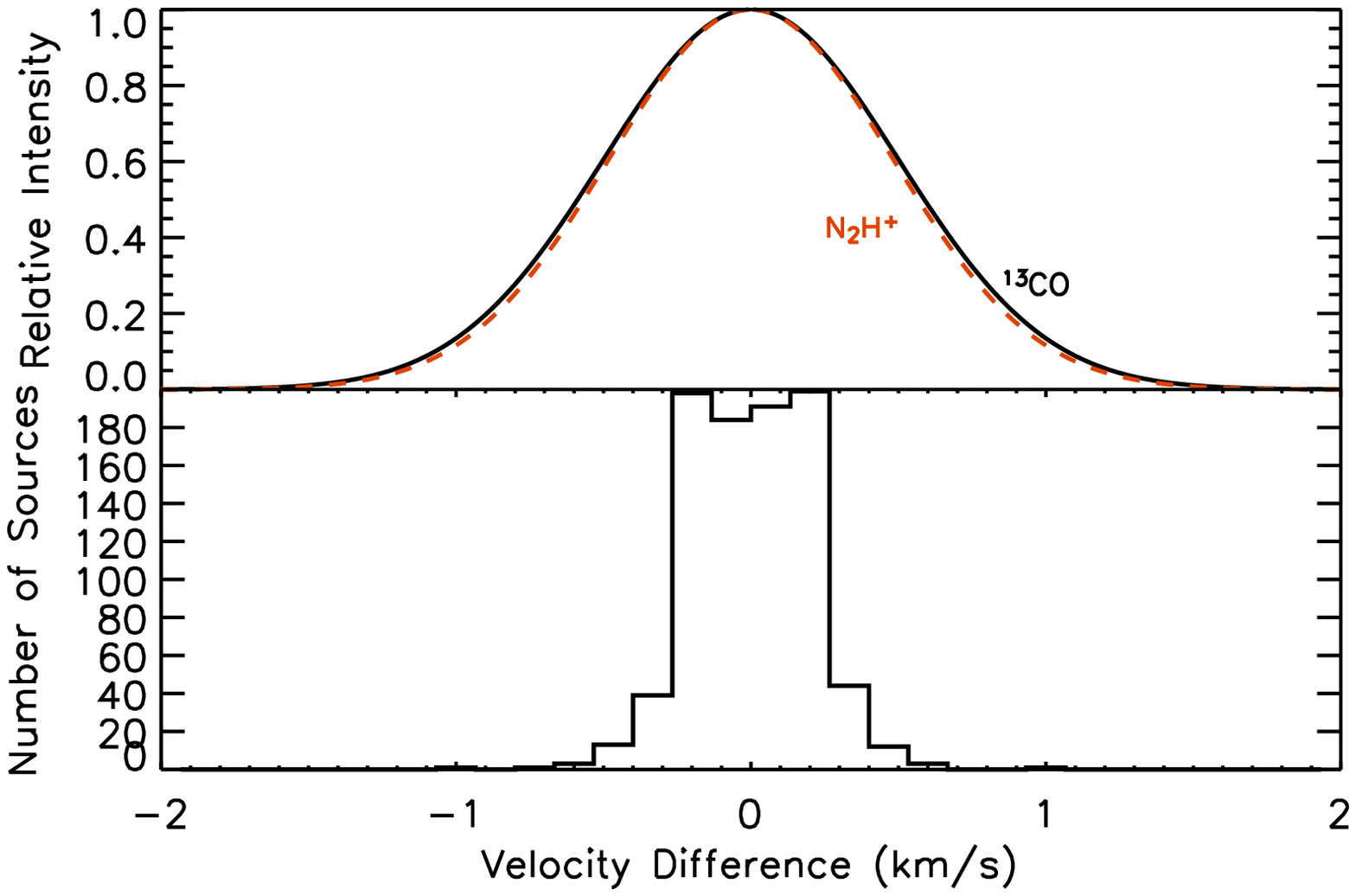}
}
\subfigure 
{
    \includegraphics[width=5.5cm]{B811.eps}
}
\subfigure 
{
    \includegraphics[width=5.5cm]{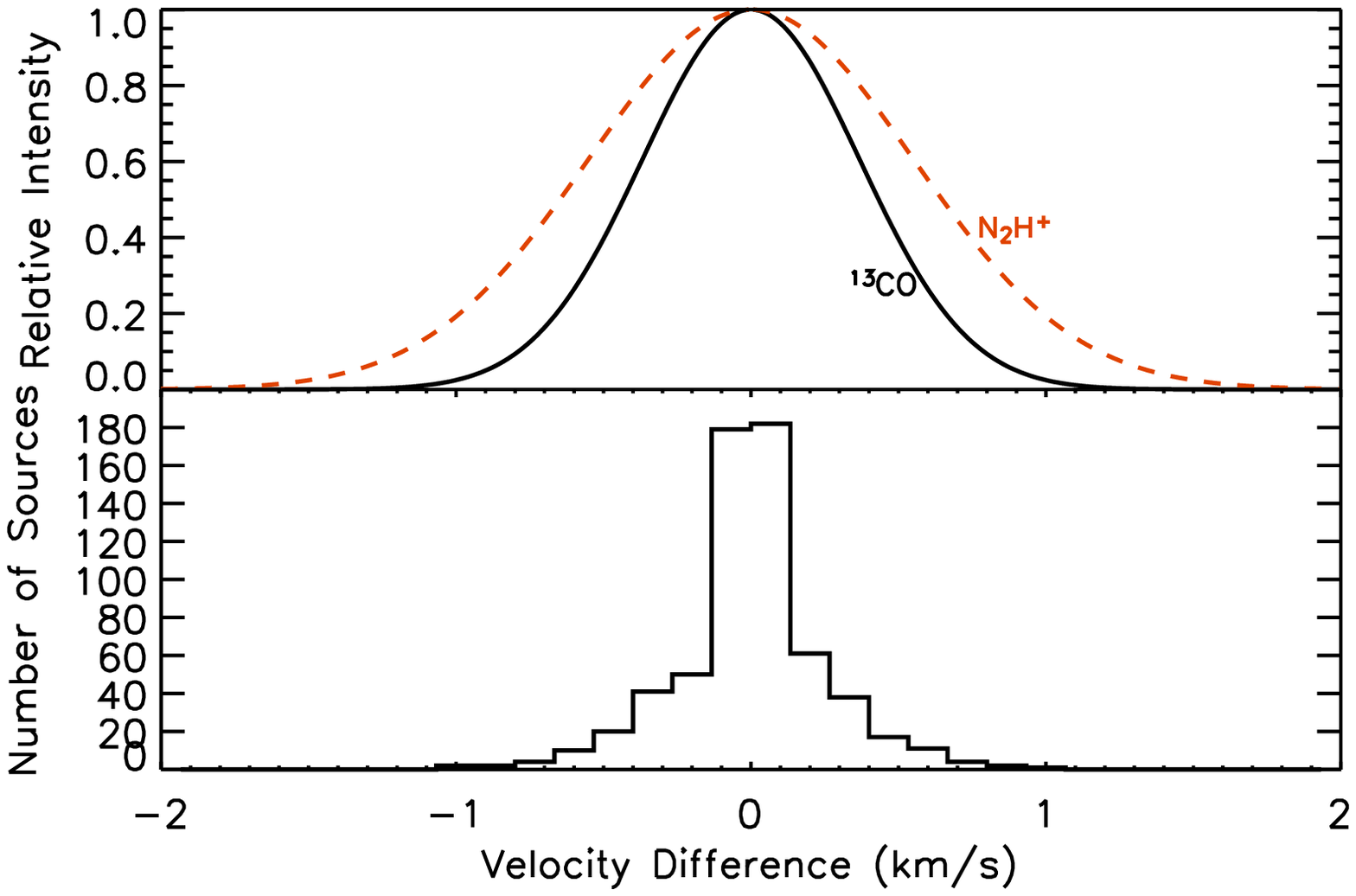}
}

\caption{\label{denshists} The effect of altering the densities used to define the high- and low-density gas.  The top panels are for the simulation at $t=1.0$ $t_{\rm ff}$, just before the first protostar forms, while the lower panel at the end of the simulation ($t=1.4~t_{\rm ff}$).  The centre panels use the standard densities of $2\times 10^5$ and $1\times 10^3$ cm$^{-3}$ to define high- and low-density gas, respectively.  In the left panels both densities are reduced by a factor of 3, while in the right panels the densities are 3 times greater.  Notice that going to higher-density tracers or later times increases the ratio of the high-density velocity dispersion (or line width) to the low-density velocity dispersion.  This can also be inferred from the left panels of Figure \ref{densplots}.  Again, with the exception of the early-time high-density tracer plots (top right), the histograms of line-centre velocity differences are significantly narrowed than either the high or low-density gas line widths.}
\end{figure*}

\section{Results}
\label{results}

After all dense core/envelope pairs had been identified in each time snapshot, viewed from all angles, and their line-centres (mean velocities) and dispersions (standard deviations) had been measured, we analysed them in the same manner as \citet{WalMyeBur2004}.  The overall results are shown in Fig.~\ref{hists} in a similar manner to those of Walsh et al.  In the top panel we plot three Gaussian distributions whose standard deviations are given by the mean standard deviations of all of the high-density (equivalent to N$_2$H$^+$) and low-density (equivalent to $^{13}$CO and C$^{18}$O) velocity spectra.  In the other three panels, we plot histograms of the distributions of differences in line-centre velocities between each pair of density tracers.  

Following Walsh et al., if there are significant motions of the cores with respect to their envelopes it is expected that the relative-velocity distributions will be similar to the width of the low-density gas (equivalent to CO) velocity spread. However, if only small random motions are observed the relative-velocity distributions will be very narrow.  Walsh et al.\ found that the spread of the relative line-centre velocity distribution was certainly smaller than the CO dispersion and probably smaller than the N$_{2}$H$^{+}$ dispersion.  We draw the same overall conclusion from the analysis of the hydrodynamical simulation: the dispersion in relative velocities is smaller than the line widths of either the low-density or high-density gas.  Thus, there is no evidence that the dense cores and envelopes have large velocities relative to one another.

Although the distributions of line-centre velocity differences are all much narrower than the gas tracer line widths, the line widths of the tracers themselves do differ from the observations.  In particular, Walsh et al.\ found that the $^{13}$CO line widths were wider than the C$^{18}$O line widths which were in turn were wider than the N$_2$H$^+$ line widths.  Averaged over our five time snaphots, we find that all three line widths are similar in the simulations -- while the $^{13}$CO line width is almost identical to that in the observations of Walsh et al., the C$^{18}$O and N$_2$H$^+$ line widths are generally wider in the simulations.  This indicates that there is a difference between the observations and simulation at high gas densities: the high-density gas velocity dispersion is usually larger in the simulations.

To investigate this, we examined the dependence of the gas velocity dispersion on density and how it evolves with time during the simulation.  In the left hand panels of Figure \ref{densplots}, we plot the gas velocity dispersion as a function of density, averaged over many view angles, for each of the five snapshots.   The arrows indicate the densities of our high (N$_2$H$^+$) and low ($^{13}$CO) density tracers.  It is clear that the velocity dispersion of the gas is a function of both density and time.  Early in the simulation, just before the first star forms, the velocity dispersion decreases almost monotonically with increasing density.  Hence the $^{13}$CO line width is larger than the N$_{2}$H$^{+}$ line width (similar to the observations).  However, later in the simulation, the velocity dispersion at high densities increases while the velocity dispersion at low densities decreases and there is a minimum gas velocity dispersion at $\sim 3\times 10^{-20}$ g~cm$^{-3}$ (i.e. $\sim 10^4$ cm$^{-3}$).  Thus, at later times, the N$_{2}$H$^{+}$ line width becomes larger than the $^{13}$CO line width.  We attribute this evolution to two effects.  In the simulation the turbulent motions in the cloud decay with time -- the turbulence is not driven.  This explains the decrease of the low-density gas velocity dispersion with time.  At high densities, as time progresses, groups of stars and brown dwarfs form in the cores.  The dynamical motions of the stars and brown dwarfs stir up the gas, increasing the velocity dispersion of the high-density gas with time.  Note that there is no obvious dependence of the line-centre velocity difference histograms on the evolutionary time of the simulation, and in all cases they are significantly narrower than the line widths.

As pointed out by Walsh et al., the main difference between their observations and the simulations is that the they observed relatively isolated low-mass cores, while the simulations to date have concentrated on star cluster formation.  Therefore, Walsh et al.'s observations are most similar to the simulation analysed at early times, when only a few objects have formed and there are no large groups of stars and brown dwarfs embedded in the dense cores.  At early times (e.g., top row of panels in Figure \ref{densplots}), we see that this is indeed when the line widths of N$_2$H$^+$ and $^{13}$CO are most similar to Walsh et al.'s observations in the sense that the line width of the high-density gas is narrower than that of the low-density gas.

In Figure \ref{denshists}, we also investigate the dependence of the results on the density of the gas being traced.  The centre panels give the results for the first and last (top and bottom) time snapshots at the standard tracer densities (i.e. $10^3$ and $2\times 10^5$ cm$^{-3}$ for $^{13}$CO and N$_2$H$^+$, respectively).  The left hand panels give the results when both densities are reduced by a factor of 3, while the right hand panels give the results when both densities are increased by a factor of 3.  What can be seen is that as the tracer densities are decreased, the line widths again are in better agreement with the observations in the sense that the ratio of the line widths of the low-density gas to the high-density gas is larger.  Thus, it may be that we have assumed that the densities which are being traced by the molecular lines are slightly too high -- reducing them by only a factor of 3 can produce a ratio of the $^{13}$CO to N$_2$H$^+$ line widths that is very similar (Fig.\ \ref{denshists}, top-left panel) to that found by Walsh et al. Whether this is correct or not could only be tested properly by including chemistry and optical depth effects in the simulations (well beyond the scope of this study).  Note that at early times, (top row in Figure \ref{denshists}) the widths of the line-centre velocity difference histograms appear to increase as the tracer densities are increased.  This does not appear to be the case later in the simulation, however.

Finally, in the right hand panels of Figure \ref{densplots}, we give the results when the simulation has been smoothed to mimic the spatial resolution of the observations.  Comparing these to the centre panels, which have not been smoothed, we see that the line widths of the N$_2$H$^+$ and $^{13}$CO are essentially independent of the smoothing.  However, in all cases the distributions of line-centre velocity differences are {\it narrower} when the analysis is performed with smoothing.  In particular, the wings of the histograms tend to disappear when smoothing is applied.  This indicates that Walsh et al.'s observations may slightly underestimate the dispersion in the line-centre velocity differences.  It also shows that our conclusion that the line-centre velocity differences are much narrower than the tracer line widths is robust against spatial smoothing.

\section{Conclusions}
\label{conclusions}

\citet{WalMyeBur2004} investigated whether or not the relative line-centre velocities of dense cores and their associated low-density envelopes in low-mass star-forming clouds were large compared to the intrinsic line widths of the cores and envelopes.  They found small relative velocities ($\sim 0.1$ km~s$^{-1}$) and interpreted this as evidence against a dynamical picture of star formation in the sense that dense cores do not appear to gain significant mass by accreting gas as they move through the lower-density environment.  In this paper, we have analysed the hydrodynamical simulation of \citet{BatBonBro2003} in a similar manner to Walsh et al.   Despite the fact that the dynamical processes of competitive accretion and dynamical interactions between stars and brown dwarfs are crucial for determining the stellar properties in this simulation, we find good agreement with the observations of Walsh et al., in the sense that the simulation also displays small relative velocity differences between dense cores and their associated envelopes ($\approx 0.2$ km~s$^{-1}$).  Thus, the dynamical picture of star formation provided by this and other similar hydrodynamical simulations is not invalidated by the observations of Walsh et al.

Our simulations do display some differences when compared with the observations in terms of the intrinsic line widths of the high- and low-density gas.  Whereas Walsh et al.\ find that the line widths of the high-density tracer are smaller than those of low-density tracers, we find that in the simulation the ratio of the line widths of high- to low-density gas is a function both of time and of the absolute density of the tracers.  There are three main effects.  First, because the simulations assume decaying rather than driven turbulence, the line width of low-density gas decreases with time.  Second, the line width of high-density gas increases with time.  We attribute this to stirring of the high-density gas as small groups of stars form within the dense cores.  Finally, if the density {\it ratio} of the high- and low-density tracers is maintained but the {\it absolute values} of the tracer densities are decreased we find that the ratio of the low-density line width to the high-density line width increases.  In general, the line width behaviour found in the simulation is most similar to Walsh et al.'s observations at early times (before the dense cores contain groups of stars) and for lower tracer densities than the canonical numbers we have used here.  We attribute the former effect to the fact that Walsh et al.'s observations were of relatively isolated low-mass cores each forming only a few stars, while the simulation analysed here is of a high-density cloud that forms groups of stars in the high-density cores that stir up the gas.  The latter effect may mean that agreement would be improved by including chemical evolution in the calculations and optical depth effects in the synthetic `observations' to better trace which gas is actually contributing to the molecular line emission.  Performing a star formation simulation of a lower density cloud with chemical evolution to test whether such a simulation does indeed give better agreement with the observations gives us a goal to aim for in the future!

\section*{Acknowledgments}

BAA and JCL thank Stuart Whitehouse and Chris Reeves for many helpful conversations. We also thank Jennifer Hatchell and Phil Myers for comments on drafts of the manuscript and the referee, Gary Fuller, for forcing us to perform a more detailed analysis than appeared in the original manuscript.  The hydrodynamical simulation analysed here was performed using the UK Astrophysical Fluids Facility (UKAFF). 
HSC was supported by PPARC standard grant PPA/G/S/2001/00515.
MRB is grateful for the support of a Philip Leverhulme Prize.

\bibliography{paper}

\end{document}